\documentclass[aps,prl,twocolumn,superscriptaddress,showkeys,amsfonts,amssymb,amsmath]{revtex4-1}
\usepackage{amsmath,latexsym,amsfonts,amssymb}
\usepackage{epsfig}
\usepackage{graphicx}
\usepackage{inputenc}
\usepackage{color}
\usepackage{ulem}
\usepackage{setspace}
\usepackage{multirow}
\usepackage{array}
\usepackage{tabularx}
\usepackage{bm}
\usepackage{color}
\usepackage[colorlinks=true,linkcolor=blue,citecolor=blue,urlcolor=blue]{hyperref}

\newcommand{\be}{\begin{equation}}
\newcommand{\ee}{\end{equation}}
\newcommand{\ben}{$$}
\newcommand{\een}{$$}
\newcommand{\ba}{\begin{eqnarray}}
\newcommand{\ea}{\end{eqnarray}}
\newcommand{\ban}{\begin{eqnarray*}}
\newcommand{\ean}{\end{eqnarray*}}
\newcommand{\bi}{\begin{itemize}}
\newcommand{\ei}{\end{itemize}}

\begin{document}

\title{Interplay between spin dynamics and crystal field in multiferroic compound HoMnO$_3$}

\author{X. Fabr\`eges}
\affiliation{Laboratoire L\'eon Brillouin, CEA, CNRS, Universit\'e Paris-Saclay, CE-Saclay, F-91191 Gif sur Yvette, France}
\author{S. Petit}
\affiliation{Laboratoire L\'eon Brillouin, CEA, CNRS, Universit\'e Paris-Saclay, CE-Saclay, F-91191 Gif sur Yvette, France}
\author{J.-B. Brubach}
\affiliation{Synchrotron SOLEIL, L'Orme des Merisiers Saint-Aubin, 91192 Gif-sur-Yvette, France}
\author{P. Roy}
\affiliation{Synchrotron SOLEIL, L'Orme des Merisiers Saint-Aubin, 91192 Gif-sur-Yvette, France}
\author{M. Deutsch}
\altaffiliation{Current address: Universit\'e de Lorraine, CNRS, CRM2, F-54000 Nancy, France}
\affiliation{Synchrotron SOLEIL, L'Orme des Merisiers Saint-Aubin, 91192 Gif-sur-Yvette, France}
\author{A. Ivanov}
\affiliation{Institut Laue-Langevin, 71 avenue des Martyrs CS 20156, 38042 Grenoble Cedex 9 - France.}
\author{L. Pinsard-Gaudart}
\affiliation{SP2M-ICMMO, UMR-CNRS 8182, Universit\'e Paris-Sud, Universit\'e Paris-Saclay, 91405 Orsay, France}
\author{V. Simonet}
\affiliation{Universit\'e Grenoble Alpes, CNRS, Institut N\'eel, 38000 Grenoble, France}
\author{R. Ballou}
\affiliation{Universit\'e Grenoble Alpes, CNRS, Institut N\'eel, 38000 Grenoble, France}
\author{S. de Brion}
\email{sophie.debrion@neel.cnrs.fr}
\affiliation{Universit\'e Grenoble Alpes, CNRS, Institut N\'eel, 38000 Grenoble, France}

\date{\today}

\newcommand{\rmn}{RMnO$_3$}
\newcommand{\ymn}{YMnO$_3$}
\newcommand{\ermn}{ErMnO$_3$}
\newcommand{\ybmn}{YbMnO$_3$}
\newcommand{\homn}{HoMnO$_3$}
\newcommand{\tc}{T$_{C}$}
\newcommand{\tn}{T$_{N}$}

\begin{abstract}
{In the multiferroic hexagonal manganite HoMnO$_3$, inelastic neutron scattering and synchrotron based THz spectroscopy have been used to investigate the spin waves associated to the Mn order together with Ho crystal field excitations. While the Mn order sets in first below 80 K, a spin reorientation occurs below 37 K, a rare feature in the rare earth manganites. We show that several Ho crystal field excitations are present in the same energy range as the magnons, and that they are all affected by the spin reorientation. Moreover, several anomalous features are observed in the excitations at low temperature. Our analysis and calculations for the Mn spin waves and Ho crystal field excitations support Mn-Ho coupling mechanisms as well as coupling to the lattice affecting the dynamics.}
\end{abstract}


\maketitle

\section{Introduction}

Multiferroics continue to attract much attention in condensed matter physics, because of the promising applications in spintronics they offer, but also because these materials raise fundamental questions. One of these concerns the understanding of the microscopic origin of the coupling between magnetic and electric degrees of freedom, and how new excitations resulting from this coupling, for instance electromagnons, can emerge. Electromagnons were first reported in 2006 as magnetic excitations excitable by the electric field component of the electromagnetic wave \cite{Pimenov2006}. Since then, electromagnons have been observed in several compounds with non colinear magnetic order such as perovskites \cite{Pimenov2006,Sushkov2007}, hexaferrites \cite{Kida2009} or even the more simple cupric oxide \cite{Jones2014}. Various mechanisms at the origin of these excitations were invoked and identified, among which the inverse Dzyaloshinskii-Moriya interaction \cite{Katsura2007}, symmetric exchange-striction mechanism \cite{Mochizuki2010b} or higher harmonics of the magnetic cycloids \cite{Mochizuki2010}. Other types of excitations involving electric and magnetic degrees of freedom have also been reported such as atomic rotation mode excited by the magnetic field component of the terahertz (THz) light \cite{Chaix2013} or hybrid excitations arising from the interplay between rare-earth crystal field (CF) excitations and 3d transition metal magnons \cite{Sirenko2008,Kang2010}, in particular in the hexagonal manganite ErMnO$_3$ \cite{Chaix2014}. All these hybrid excitations often lie in the THz range and are also investigated using inelastic neutron \cite{Pailhes2009} or X ray scattering \cite{Toth2016} and Raman spectroscopy \cite{Aupiais2018}.

\begin{figure}[!h]
\includegraphics[width=8cm]{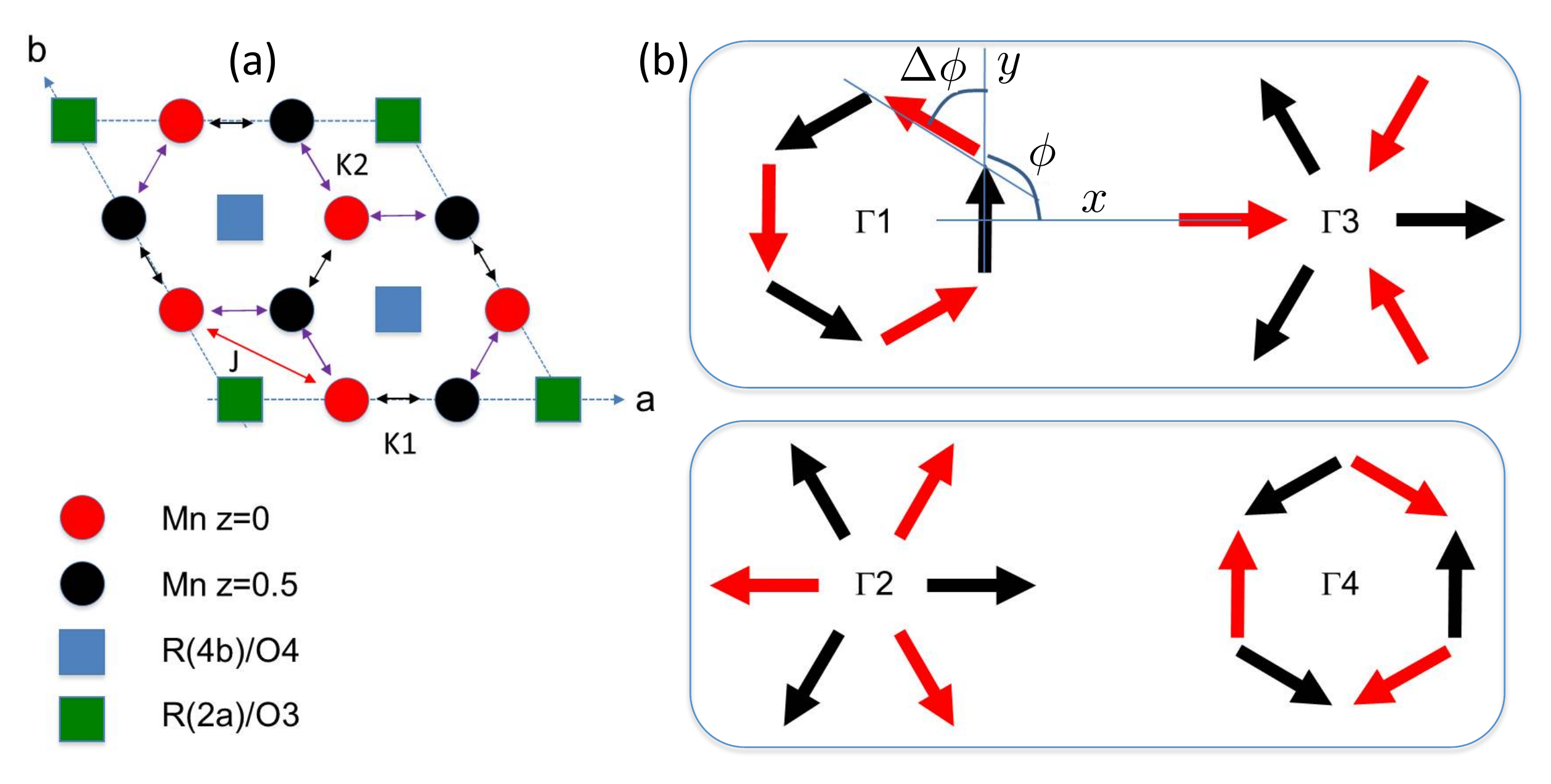}
\caption{Sketch of the \homn\ crystallographic and magnetic structures. (a) Projection of the different atomic positions onto the ab-plane, omitting the oxygens for clarity. The Mn ions on the 6c Wyckoff site have a $x$ coordinate within the ab-plane close to 1/3 and a $z$ coordinate along the c-axis equal to 0 (red spheres) or 1/2 (black spheres). The rare-earths occupy two different Wyckoff sites (2a and 4b). The main intra-plane ($J$) and inter-layer ($K_{1,2}$) exchange interactions between the Mn magnetic moments are indicated. (b) Magnetic arrangement of the Mn ions with zero propagation vector corresponding to the four one-dimensional irreducible representations $\Gamma_i$ of the P6$_3$cm space groupe. The homometric pairs, indistinguishable through neutron scattering if $x_{Mn}$=1/3, are indicated by the boxes.}
\label{fig1}
\end{figure}

The hexagonal manganites h-\rmn\, where R= Y, Er, Ho, etc, crystallizing in the P6$_3$cm space group, have been widely studied \cite{Bertaut1963,Katsufuj2002,Cheong2007,Sim2016} since they combine remarkable properties associated to multiferroism and magnetic frustration. They belong to type I multiferroics, where ferroelectricity and magnetic orders occur at different temperatures: onset of an electric polarization along the c-axis at much higher temperature (\tc$\simeq$800K) than the ordering of the Mn magnetic moments (\tn$\simeq$80K). Despite these different energy scales, non negligible coupling between the electric and magnetic degrees of freedom are present. For instance, the Mn magnetic ordering is accompanied by ionic motions inside the unit cell \cite{Fabreges2009,Lee2008} giving rise to a giant magneto-elastic coupling, connected with an increase of the ferroelectric polarization. The Mn sublattice consists of triangular planes (Fig. \ref{fig1}) with a peculiar stacking along the c-axis, each Mn$^{3+}$ ion sitting close to the middle of the triangle of Mn$^{3+}$ belonging to adjacent layers. The actual materials have then in common a 120$^{\circ}$ spin structure for the Mn magnetic moments in the triangular planes, but the stacking and relative phase difference of these patterns depend on the R atom ((magnetic R=Ho, Yb, or not R= Y, Sc), and correspond to the four one-dimensional irreducible representations shown in Fig. \ref{fig1}. A key parameter tuning the sign of the exchange interaction and the Mn$^{3+}$ single-ion anisotropy seems to be the shift away from 1/3 of the $x$ position of the Mn$^{3+}$ ions within the triangular plane  \cite{Fabreges2009,Solovyev2014}. The ions R$^{3+}$ are situated on two Wyckoff sites 2a and 4b. For magnetic rare earths, they also order at lower temperature, either polarized by the Mn molecular field (site 4b) or from their mutual interaction in a temperature range of few K (site 2a) \cite{Fabreges2008,Chattopadhyay2018}.

In this study, we focus on the Ho compound, which has been intensively studied for its rich $H-T$ phase diagram and its strong magnetoelectric effects \cite{Cheong2007,Lottermoser2004}. It becomes ferroelectric at 875 K and the Mn magnetic moments order at T$_N$=75 K in the $\Gamma_4$ representation (see Fig. \ref{fig1}). Remarkably, a spin reorientation occurs at T$_{SR}$=37 K which is concomitant with the ordering of the Ho(4b) magnetic moments, while the Ho(2a) moments order around $T_{Ho}$=5 K, both along the c-axis \cite{Munoz2001,Fiebig2002,Brown2006,Brown2008,Nandi2008,Fabreges2009}. This Mn reorientation transition, unique in the hexagonal manganites, shows up also in the electric properties with a drop in the electric polarisation below T$_{SR}$ which is further recovered at $T_{Ho}$ \cite{Hur2009}.
Signatures of the rare-earth Mn coupling in the spin dynamics in an applied magnetic field have also been reported \cite{Talbayev2008,Laurita2017}, although no rare-earth CF-Mn magnon hybridization has yet been observed.

We report here a detailed study on the dynamical properties of this compound in the THz range using both electromagnetic waves and neutrons as complementary probes. We show that the THz dynamics is particularly complex in \homn\ due to the presence of several Ho CF excitations in addition to the Mn spin waves that surprisingly  become non-dispersive along the c-axis at low temperature. To interpret the observed spectra and their remarkable temperature dependence through the spin reorientation transition and below, we have performed calculations both for Mn and Ho dynamics. Our work brings new insight on the complex spin dynamics of this compound including possible signatures of the couplings between the different magnetic ions, but also with their lattice surroundings.

\section{Experimental details}

Single crystals have been grown using the floating zone method, yielding rods of about 7 mm in diameter and 4 cm long. From these crystals, two small plaquettes were cut for the THz measurements (Typically 2 mm in diameter and 300 $\mu$m in thickness): one with the c-axis perpendicular to the plaquette, the other one with the c-axis within the plaquette. The polarized THz emission of the synchrotron at SOLEIL was used to probe the THz properties similarly to the study on ErMnO$_3$ \cite{Chaix2014}.

\begin{figure*}[!t]
\includegraphics[width=16cm]{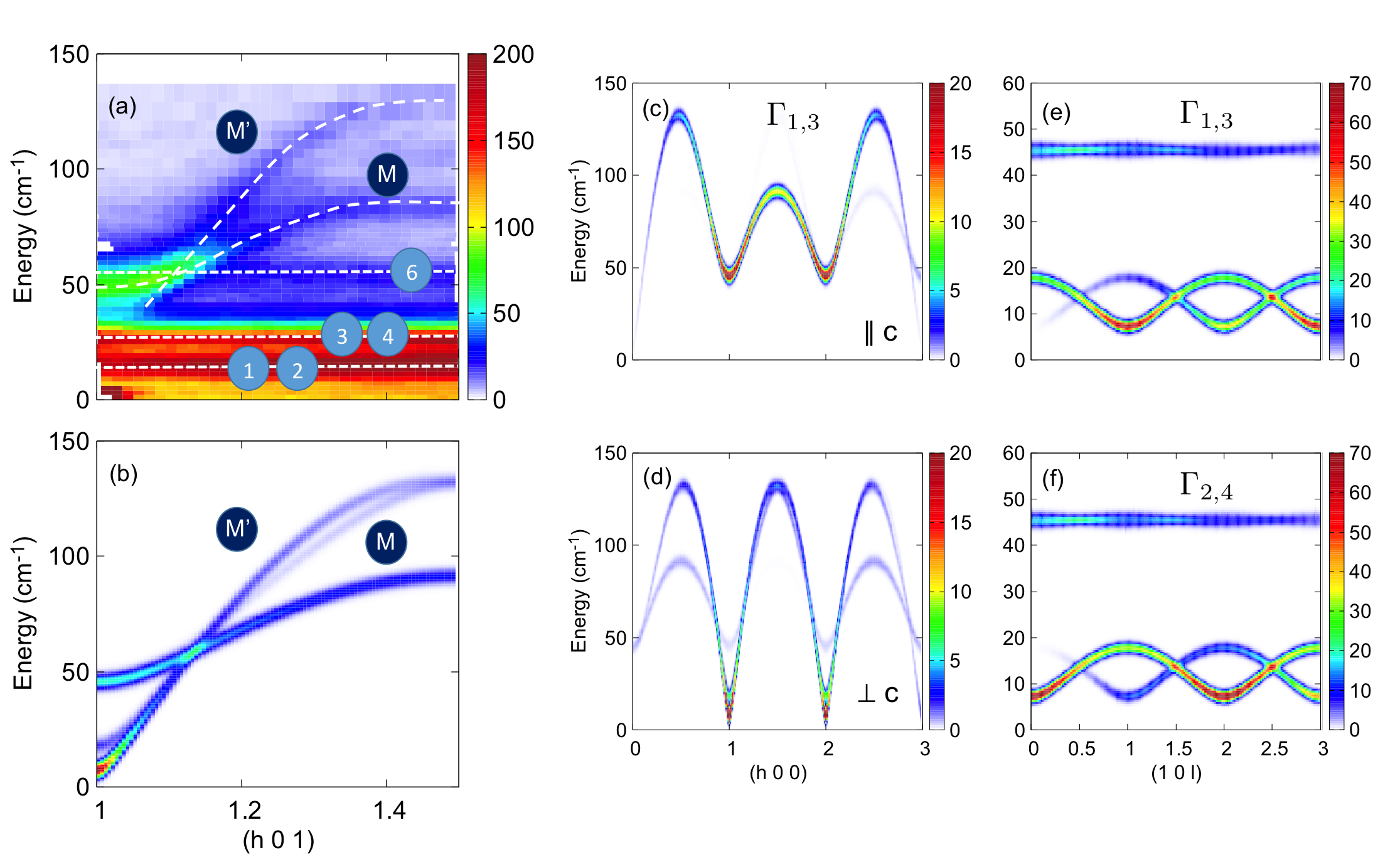}
\caption{Spin dynamics in \homn\. (a) Mn$^{3+}$ spin waves and Ho$^{3+}$ CF excitations measured by INS along the $(h, 0, 1)$ reciprocal space direction. These data have been taken on IN8 at 2K with a medium energy resolution of about 0.8 meV. M and M' denote the Mn dispersive spin waves that can be seen at the zone center. Labels 1-6 refer to non dispersive Ho CF lines. (b) Comparison with spin wave calculations using the model described in the text for the $\Gamma_{1,3}$ configurations of the Mn magnetic structure. (c,d) Spin wave calculations along $(h, 0, 0)$ in the $\Gamma_{1,3}$  configurations for spin fluctuations respectively along the c-axis and within the basal hexagonal plane. (e-h) Spin wave calculations along $(1, 0, \ell)$ for the 4 different possible $\Gamma_i$. In the calculations of panels (b-f), modest but finite values of $D_x$ and $D_y$ were taken into account to ensure the stability of the various $\Gamma_i$, see Table \ref{table-gamma}. This is responsible for the rise of the small gaps at zone centers. All the spin wave calculations have been performed using the SpinWave software \cite{SpinWave}.}
\label{INS1-v4}
\end{figure*}

Inelastic neutron scattering (INS) measurements were performed at the thermal spectrometer IN8 (ILL) as well as at the 4F1 and 4F2 cold triple-axes spectrometers installed at LLB. The final wave vector used was $k_f$ = 2.662, 1.55 or 1.3 \AA$^{-1}$ depending on the desired energy resolution (0.8, 0.25 and 0.12 meV respectively). Higher order reflections were removed using a PG (IN8) or nitrogen cooled Beryllium filter (4F). The sample was mounted to have access to $Q=(h,0,\ell)$ scattering wave-vectors and attached to the cold finger of an orange cryostat. Several Energy-scans taken along $Q=(h,0,\ell)$ were collected to construct the maps shown in Fig. \ref{INS1-v4} and \ref{INS2-v4}.

\section{Inelastic neutron scattering}

\begin{figure*}[!t]
\includegraphics[width=16cm]{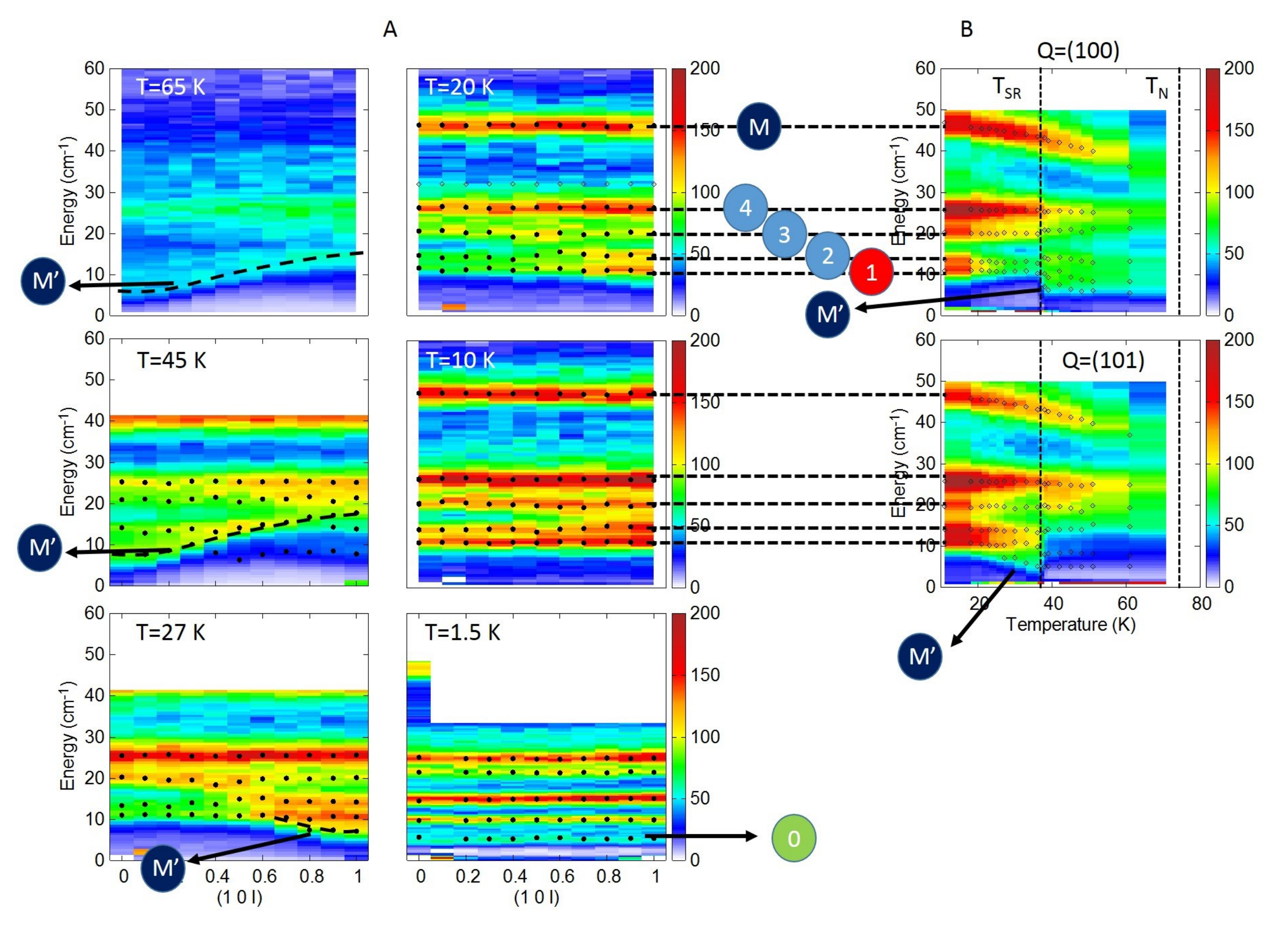}
\caption{Panel A: Inelastic neutron scattering color maps in \homn\ for the $(1, 0, \ell)$ reciprocal space direction at different temperatures, ranging from 65 K down to 1.5 K. The spin reorientation occurs at T$_{SR}=$37 K. Dispersing spin waves branches (labeled M and M')a ssociated to Mn order can be observed at 65, 45 and 27 K. The spectral weight is typical of the $\Gamma_{2,4}$ configuration at 65 K and 45 K, but it becomes typical of $\Gamma_{1,3}$ at 27 K. In addition, flat modes (0-4), likely Ho crystal field excitations, are also observed. For the temperatures below 27 K, the spin waves gradually fade away into the CF excitations and only non-dispersive excitations remain at 1.5 K (below T$_{Ho}$). The dots added on the color maps were obtained by gaussian fits of the individual constant-$Q$ E-scans.
Panel B : inelastic neutron scattering color maps in \homn\ for the $(100)$ and $(101)$ reciprocal space positions as a function of temperature. The dots added on the color maps were obtained by gaussian fits of the individual constant-$Q$ E-scans. The dashed vertical lines indicate the transition temperatures.
\label{INS2-v4}}
\end{figure*}

The magnetic excitations have been first investigated by INS. Figures \ref{INS1-v4}a and \ref{INS2-v4} give an overview of those results, showing the spectra along ${\bf a^*}$, along ${\bf c^*}$, and the temperature dependence measured at the zone centers (1, 0, 0) and (1, 0, 1). Dispersive spin waves are clearly visible along ${\bf a^*}$ with a maximum of the branches around 17 meV (140 cm$^{-1}$) while the amplitude of dispersion along ${\bf c^*}$ is much weaker, reflecting the hierarchy of the magnetic interactions much stronger within the ab-planes than in between. In addition to the spin waves, non-dispersive excitations are visible corresponding to transition between CF levels of the two Ho$^{3+}$ ions on 4b and 2a sites. Remarkably, a significant temperature dependence of the magnetic excitations is visible at T$_{SR}$ as shown at the zone centers (1, 0, 0) and (1, 0, 1) in Fig.\ref{INS2-v4} where changes of spectral weight are observed for all the excitations below 30 cm$^{-1}$. More surprisingly, along ${\bf c^*}$, the spin waves dispersion disappears gradually below T$_{SR}$ and finally the corresponding branches remain flat below $\thickapprox$ 10 K.

The coexistence of both spin waves and CF transitions, along with a subtle temperature evolution, make however the interpretation of those spectra quite complex. Fortunately, previous studies \cite{Sato2003,Park2003,Vajk2005,Petit2007,Fabreges2008,Fabreges2009,Pailhes2009,Kim2018} performed for several members of the \rmn\ provide valuable assistance. They especially concluded to the relevance of the following Hamiltonian to describe the Mn spin waves:
\begin{eqnarray}
{\cal H} &=& \frac{1}{2}\sum_{p, \langle i,j \rangle} J_{i,j} {\bf S}_{i,p} {\bf S}_{j,p} + \frac{1}{2}\sum_{\langle i,j \rangle} K_{i,j} {\bf S}_{i,1} {\bf S}_{j,2} \nonumber\\
& & + \sum_{i,p} D_z ({\bf S}_{i,p}^z)^2 - D_x ({\bf S}_{i,p}^x)^2 - D_y ({\bf S}_{i,p}^y)^2
\label{H3d}
\end{eqnarray}
In this expression, ${\bf S}_{i,p}$ is the Mn spin at site $i$, $p=1, 2$ is an index that labels the two Mn planes in the unit cell, $D_z$ is an easy plane anisotropy ($D_z \ge0$), and $D_{x,y} \ge0$ are effective easy axis anisotropies, forcing the spins to point along the $x$ or $y$ directions (see Fig.\ref{fig1}). $J_{i,j}$ is the antiferromagnetic coupling between neighboring spins in the same plane, and $K_{i,j}$ the antiferromagnetic coupling between neighboring spins in adjacent planes. As shown in Fig. \ref{fig1}a, the $x_{Mn}$ position allows to distinguish two different paths for $K$, a long and a short one, labeled hereafter as $K_{1}$ and $K_{2}$. To proceed, we first assume $D_x=0$ and $D_y=0$. Straightforward classical minimization of ${\cal H}$ shows that the ground state structures are planar 120$^{\circ}$ spin configurations. Using the convention of Figure \ref{fig1}, the moments are given by :
\begin{equation}
{\bf S}_{i,p} =
S \left(
\begin{array}{c}
\cos (\phi_p + (i-1) \frac{2\pi}{3}) \\
 \sin (\phi_p + (i-1) \frac{2\pi}{3}) \\
 0
\end{array}
\right)
\end{equation}
and the classical energy writes:
\begin{equation}
{\cal E} = -3JS^2 + 2(K_{1}-K_{2}) S^2 \cos \Delta \phi, ~\mbox{with}~\Delta \phi=\phi_{2}-\phi_1
\end{equation}
The value of $\phi_1$ along with the relative phase difference $\Delta \phi=\phi_{2}-\phi_1$ between two adjacent planes fully determine the final structure. These structures can be classified using the $\Gamma_{i=1,2,3,4}$ irreducible representations listed in Table \ref{table-gamma}. Note that two 2-dimensional irreducible representations are also allowed in the P6$_3$cm space group but were not considered in this work. The sign of $K_{1}-K_{2}$ determines the stability of each phase: $\Gamma_{1,2}$ are stabilized for $K_{1}-K_{2} \ge 0$, while $\Gamma_{3,4}$ are stabilized for $K_{1}-K_{2} \le 0$.

\begin{table}
\begin{tabular}{p{1cm}|p{1cm}|p{1cm}|p{5cm} }
$\Gamma_i$ & $\phi$ & $\Delta \phi$ & Conditions\\
\hline
1 & $\pi/2$ & $\pi$ & $D_y \ge D_x, K_1-K_2 \ge 0$ \\
2 & 0 & $\pi$ & $D_y \le D_x, K_1-K_2 \ge 0$ \\
3 & 0 & 0 & $D_y \le D_x, K_1-K_2 \le 0$ \\
4 & $\pi/2$ & 0 & $D_y \ge D_x, K_1-K_2 \le 0$ \\
\end{tabular}
\caption{Condition for the Mn$^{3+}$ single-ion anisotropy and the interplane couplings stabilizing one 3 dimensional  magnetic order of the Mn out of the four irreducible representations $\Gamma_i$.}
\label{table-gamma}
\end{table}

In principle, six spin wave branches are expected in this model. However, since $K_{1}\approx K_{2}$, these branches should be considered as three almost doubly degenerate modes. Moreover, due to the hexagonal symmetry, these three branches can be described by a single one, stemming from $(0, 0, \ell)$ and going soft at $(3,0,\ell)$. The two other branches are identical but stem from $(1, 0, \ell)$ and $(2,0,\ell)$ and go soft respectively at $(4,0,\ell)$ and $(5,0,\ell)$. As shown in Fig.$\ref{INS1-v4}$(c,d), spin wave calculations indicate that the former corresponds mostly to correlations between spin components along the c-axis (Fig.$\ref{INS1-v4}$c), while the two latter correspond mostly to in-plane correlations (Fig.$\ref{INS1-v4}$d). As a result, two acoustic-like and one optical-like spin wave emerge from each zone center. The gap $\Delta$ of the latter occurs at around 40 cm$^{-1}$ (5 meV) and is related to the easy-plane anisotropy by $\Delta \approx 3S\sqrt{D_z J}$.

As explained above, the degeneracy is lifted as soon as $K_{1} \ne K_{2}$. This becomes especially clear when looking at the dispersion along $(1, 0, \ell)$. This point is illustrated in Fig.\ref{INS1-v4}e-f, showing the calculated spin wave dispersions for the different $\Gamma_i$ magnetic configurations. We notice that the spectral weight of the lowest energy mode is strong at $(1,0,1)$ for $\Gamma_{1,3}$, and at $(1,0,0)$ in the case of $\Gamma_{2,4}$. This change has been put forward in Ref. \onlinecite{Fabreges2009} to explain that the spin reorientation process in \homn\ involves a symmetry change from $\Gamma_{2,4}$ below T$_N$ to $\Gamma_{1,3}$ below T$_{SR}$ with a 180$^{\circ}$ rotation of the spins in one of the triangular layer. This change was attributed to a change of sign of $K_{1}-K_{2}$ induced by the change in position of the Mn ions from $x_{Mn}=0.325$ above T$_{SR}$ to $x_{Mn}\approx0.334$ below. In Ref. \onlinecite{Fabreges2009}, the authors assumed $\Gamma_1$ to be the ground state. Actually, non linear optical measurements as well as other probes suggest rather a $\Gamma_3$ magnetic order below T$_{SR}$ \cite{Fiebig2002,Brown2006,Nandi2008}. Since the $\Gamma_{1,3}$ and $\Gamma_{2,4}$ pairs are homometric, i.e. they cannot be distinguished by unpolarized neutron scattering if $x_{Mn}$ is strictly equal to 1/3, we will assume that $\Gamma_3$ is the actual ground state in the following.

With these results in hand, we come back to the analysis of the experimental results. Fig.\ref{INS1-v4}b displays the calculated inelastic neutron intensity for {\bf Q} wave vectors along $(h,0,1)$. Using $J=$2.5 meV and $D_z=$0.65 meV, the high energy behavior of the spectrum is well reproduced. For instance, the spin wave labeled M originates from the optical-like mode at $\Delta$. It is clearly visible on Figure \ref{INS1-v4}. In Fig.\ref{INS2-v4} (panel A), it looks like a flat band along ${\bf c^*}$ due to the weakness of the interplane couplings $K_1$ and $K_2$. Finally, as shown in panel B, $\Delta$ hardens towards 45 cm$^{-1}$ when decreasing temperature. Note that some renormalisation of the Mn excitations has been reported in \homn\ as due to magnon-phonon coupling \cite{Kim2018}.

The second dispersive spin wave labeled M' is the acoustic-like spin wave mode stemming from the zone center and consisting of two interlaced branches. It can be also identified in Fig.\ref{INS2-v4}, thanks to the strong intensity expected for $\Gamma_{2,4}$ structures at (1, 0, 0). Below T$_{SR}$, this spectral weight jumps to (1, 0, 1), as a result of the transition to $\Gamma_{1,3}$. From panel B in Figure \ref{INS2-v4}, the change of spectral weight from (1,0,0) to (1,0,1) at T$_{SR}$ can also be noticed. The bandwidth of M' is well captured for $K_1-K_2=0.003$ meV. The gap of this mode originates from small single-ion anisotropies of the Mn$^{3+}$ (terms $D_y$ and $D_x$). An additional origin of anisotropy is expected to come from the Ho molecular field that should lead to a scaling of the gap with the Ho(4b) ordered moment. Indeed, at the mean field level, the coupling between the 3d and 4f magnetic species writes:
\ben
{\cal V} = \sum_{m,i,p,\ell} {\bf S}_{m,i}~g_{m,i,p,\ell}~\langle {\bf J}_{p,\ell}\rangle
\label{Hcc}
\een
where $g$ is the coupling tensor, $m,p$ and $i,\ell$ are indices that label the unit cell and the atom for the 3d and rare earth sublattices. $\langle {\bf J}_{p,\ell}\rangle$ is the average magnetic moment at the rare earth site. Combined with the exchange Hamiltonian described in Eq. \ref{H3d}, ${\cal V }$ polarizes the Mn spins, hence plays the role of anisotropy. In principle, this should result in a gap in the spin wave spectrum as observed in YbMnO$_3$ \cite{Fabreges2010}. However, in the present case, the expected scaling seems not to fully operate probably due to the coupling of the Mn spin waves with Ho CF excitations.

\begin{figure*}
\resizebox{19cm}{!}{ \includegraphics{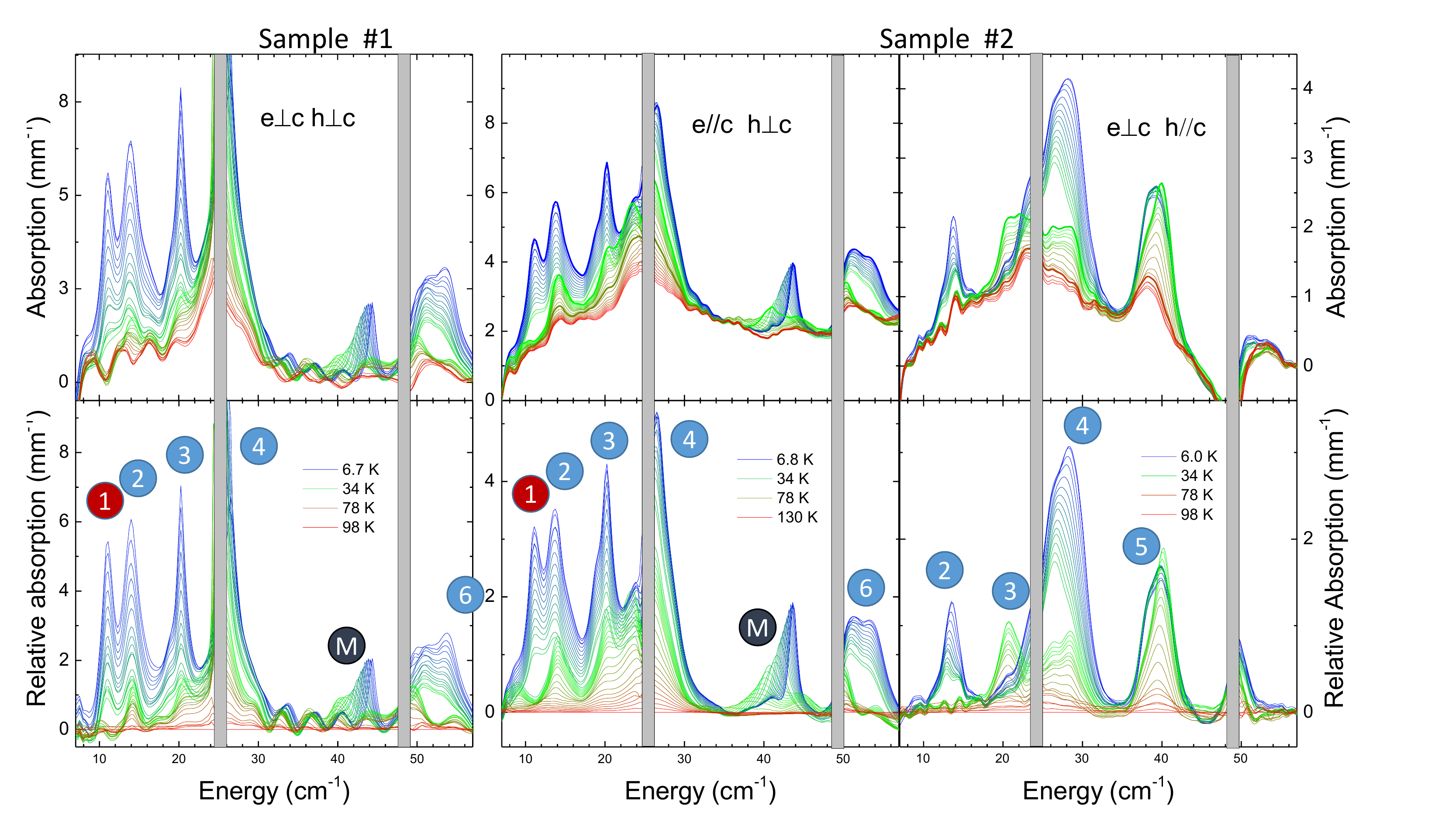}}
\caption{THz absorption spectra of HoMnO$_3$ for the 3 different orientations of the electromagnetic wave {\bf e}, {\bf h} fields with respect to the crystal c-axis, using two different samples. Upper panels: absolute absorptions. Lower panels: relative absorptions with a reference above the magnetic ordering temperature at 100 K or 130 K.}
\label{fig2}
\end{figure*}

Indeed, in addition to the spin waves, non-dispersive excitations are also visible in Fig.\ref{INS1-v4} and \ref{INS2-v4}. These features correspond to CF transitions of the Ho$^{3+}$ ions on 4b and 2a sites and are labeled 0-6. As we shall see in the next part, those labels are also used to describe the THz data (see Figure \ref{fig2}). Note that the mode 5 is not visible in our neutron scattering experiments probably because it overlaps with the magnon M or because its neutron cross section is intrinsically weak. Panel B of Figure \ref{INS2-v4} shows the temperature dependence of CF transition modes 1 to 4. The intensity of those modes gradually increases with decreasing temperature while their positions slightly change, but the most remarkable feature is the formation of an additional line labeled 1 just below T$_{SR}$. Note that the line labeled 0 appears at very low temperature only, hence probably due to the Ho ordering on site 2a. In addition, quite strikingly, the dispersion of M' eventually merges gradually below T$_{SR}$ into those CF lines (at least into those below 30 cm$^{-1}$). Finally, the spectrum encompasses only flat bands below $\thickapprox$ 10 K.

We point out that this very peculiar temperature evolution of M' shows that this branch crosses several Ho crystal field excitations, which makes very likely hybridization mechanisms (via a resonant process) between the Mn spin wave dispersion and the CF lines at low temperature.

\begin{figure}
\resizebox{8.5cm}{!}{ \includegraphics{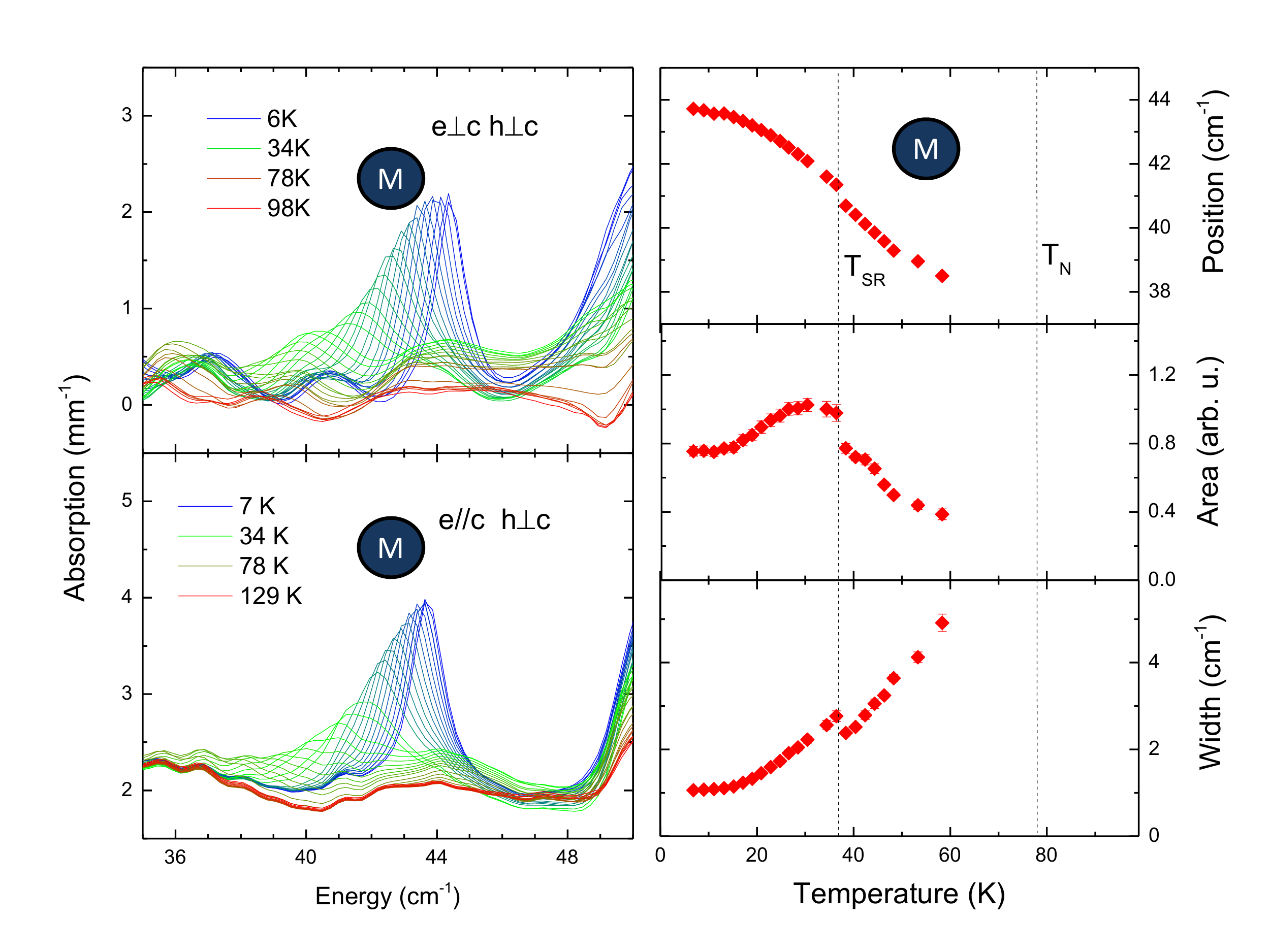}}
\caption{Details of the THz spectra around the magnon excited for a THz {\bf h} field perpendicular to the c-axis. Left: Absorption measured at different temperatures for two different direction of the THz {\bf e} field on two different samples. Right: results of the gaussian fits of the absorption peak with the position, area and width as a function of temperature obtained for ${\bf e}\parallel{\bf c}$ (with similar results for ${\bf e}\perp{\bf c}$). Note the emergence of the magnon M below T$_N$ and how it is affected by T$_{SR}$.}
\label{fig5}
\end{figure}

\begin{figure}
\resizebox{8.5cm}{!}{ \includegraphics{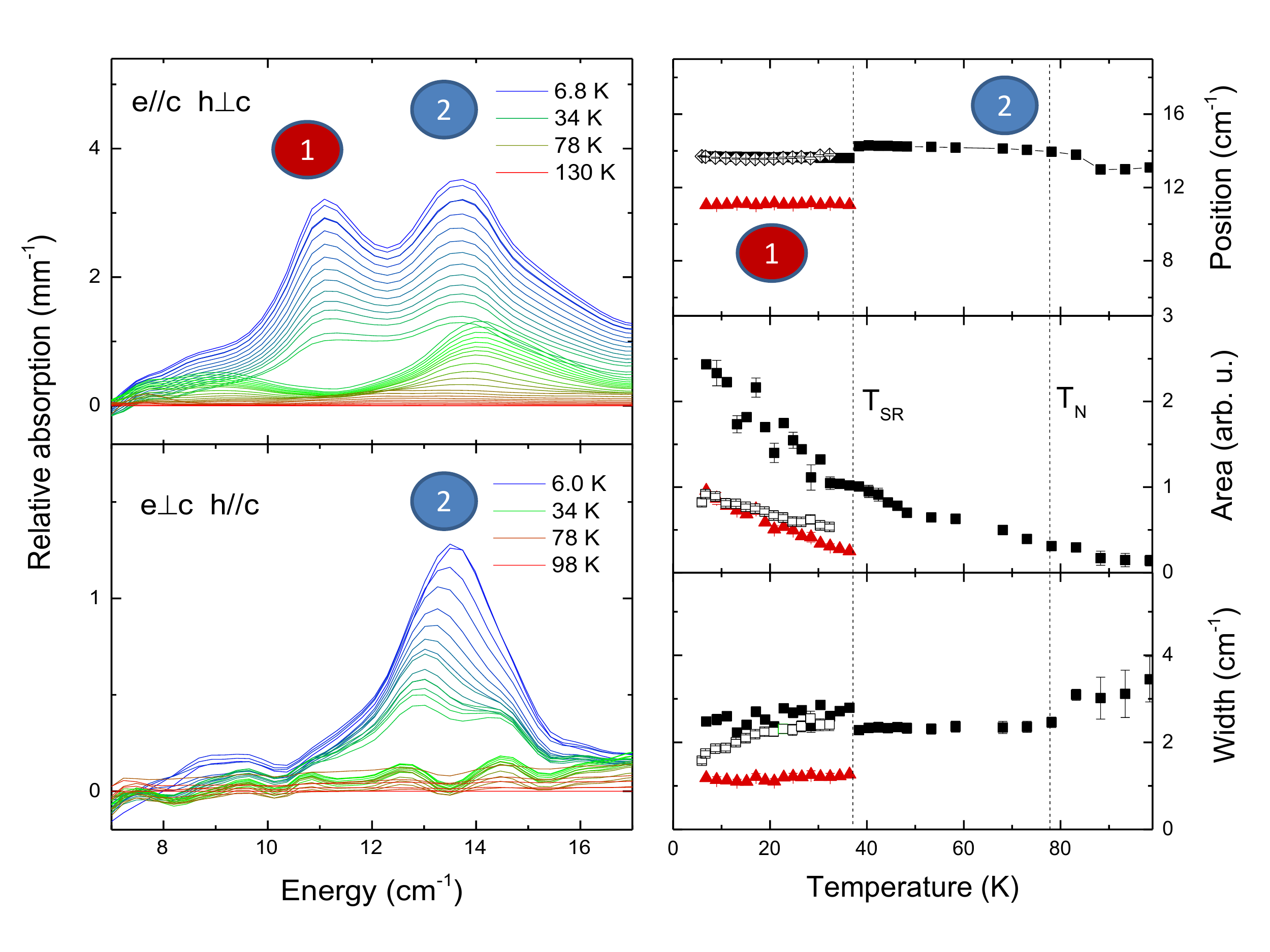}}
\caption{Details of the THz spectra around band 1 and 2. Left: relative absorption measured for two different orientations of {\bf e} and {\bf h} on the same sample. Right: results of the gaussian fits for the absorption peaks with the position, area and width as a function of temperature. Closed and open symbols refer to {\bf h$\perp$c} and ${\bf h}\parallel{\bf c}$ respectively.  Below T$_{SR}$, two new excitations appear abruptly: band 1 for {\bf h$\perp$c} and band 2 for ${\bf h}\parallel{\bf c}$.}
\label{fig3}
\end{figure}


\section{Synchrotron based THz spectroscopy}

To get a better insight into the spin dynamics, we have used THz spectroscopy as a complementary probe of both Mn spin waves and Ho Crystal field excitations. In this electromagnetic absorption technique, due to the high velocity of photons in the material, the measurements are considered in the long wave length limit and comparison with neutron data is therefore relevant for the (0, 0, 0) center of the Brillouin zone. There, all Ho crystal field excitations should be present. As for the the Mn spin waves, calculations similar to those presented in the previous section show that the four different spin configurations $\Gamma_{i}$ have the same signature at (0,0,0). It consists of  a high energy branch (M) expected around 45 cm$^{-1}$ which involves correlations between spin components perpendicular to the c-axis, and a single low energy branch M' corresponding to spin correlations along the c-axis that is pushed to finite energy when some spin anisotropy is present (see above and Fig.\ref{INS1-v4}).

The THz spectrum of HoMnO$_3$ was first studied in \cite{Talbayev2008} where the Mn (M) magnon mode as well as several Ho crystal field excitations were observed as a function of external magnetic field.  The Mn magnon was more recently investigated under light irradiation \cite{Bowlan2016} while its dependence on the Ho-Mn interaction is reported in \cite{Laurita2017}. Here, thanks to synchrotron based spectroscopy, we are able to measure with great precision the whole spectra from 0.3 to 1.8 THz (10 to 60 cm$^{-1}$).

\begin{figure}
\resizebox{8.5cm}{!}{ \includegraphics{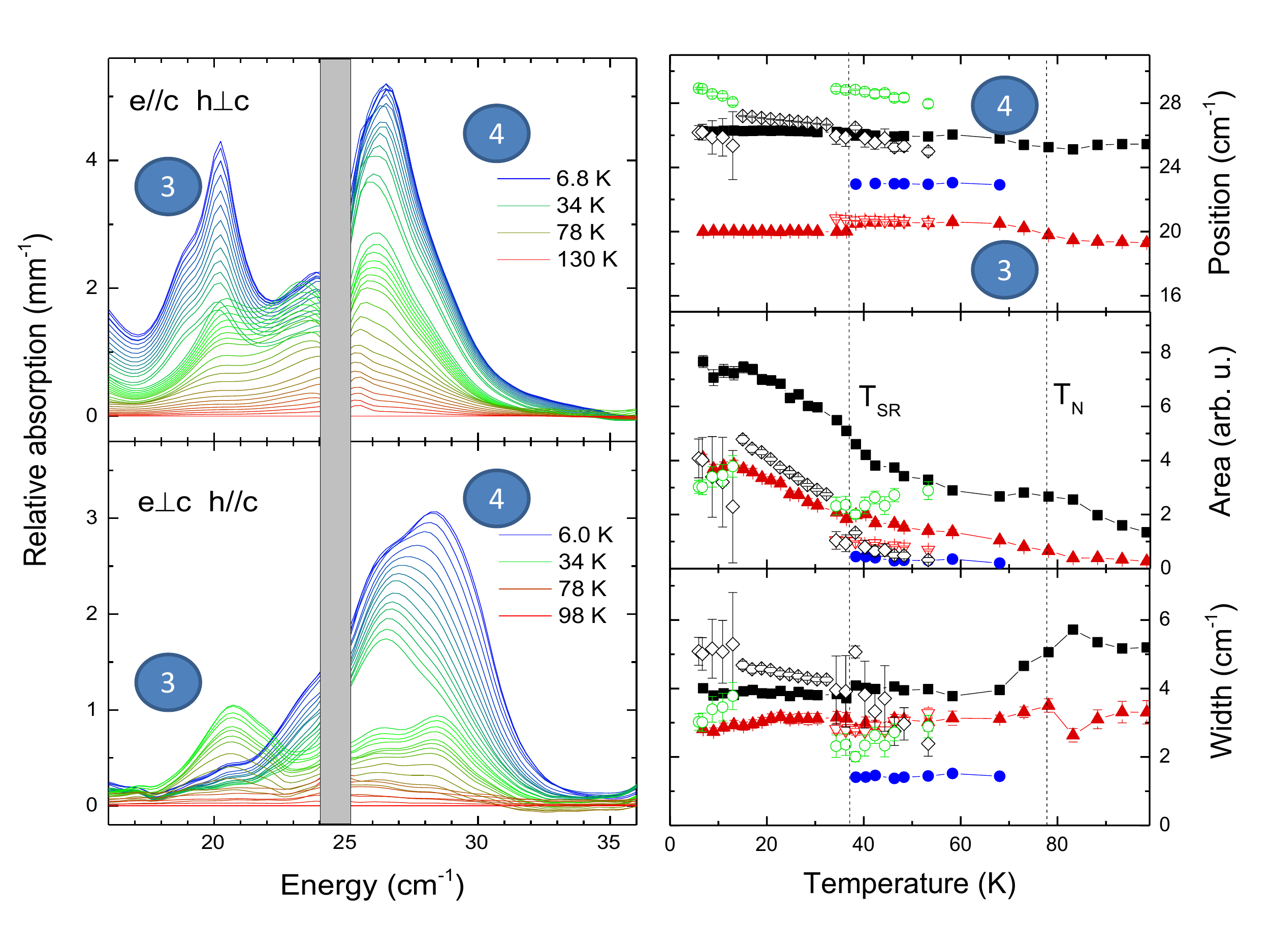}}
 \caption{Details of the THz spectra around band 3 and 4. Left: relative absorption measured for two different orientations of {\bf e} and {\bf h} on the same sample. Right: results of gaussian fits for the absorption peaks with the position, area and width as a function of temperature. A maximum of 3 peaks were used. Closed and open symbols refer to {\bf h$\perp$c} and ${\bf h}\parallel{\bf c}$ respectively. Note the complexity of the spectra with several components that are all affected by T$_{SR}$.}
 \label{fig4}
 \end{figure}

All the spectra recorded at different temperatures from 130 K or 100 K down to 6 K are shown in Fig.\ref{fig2} for the three possible orientations (as regards the c-axis) of the THz electric {\bf e} and magnetic {\bf h} fields. Several absorption bands are observed from 11 cm$^{-1}$ up to 55 cm$^{-1}$, quite differently from other hexagonal manganites such as YMnO$_3$ or ErMnO$_3$, suggesting the influence of Ho ions in these THz spectra. Further inspection shows that, for the two different THz polarizations with the same magnetic field direction {\bf h$\perp$c}, the spectra are very similar and differ substantially from the case where  ${\bf h}\parallel{\bf c}$. This is a strong indication that all the observed excitations are of magnetic origin. The Mn magnon (M) expected around 45 cm$^{-1}$ is clearly observed  for {\bf h$\perp$c} and seems to behave as expected for a conventional magnon by emerging below T$_N$. It shifts to higher energies when the temperature is lowered (see Fig.\ref{fig5}). This magnon involves spin components perpendicular to this axis, in agreement with the THz magnetic selection rule {\bf h$\perp$c}. The other Mn magnon (M') involving the spin component along the c-axis is expected at lower energies for ${\bf h}\parallel{\bf c}$ but is not easily distinguishable. All the other observed bands are specific to the Ho compound and should be related to Ho crystal field excitations. Quite surprisingly, all spectra are strongly affected by the spin reorientation process at T$_{SR}$.

 \begin{figure}
\resizebox{8.5cm}{!}{ \includegraphics{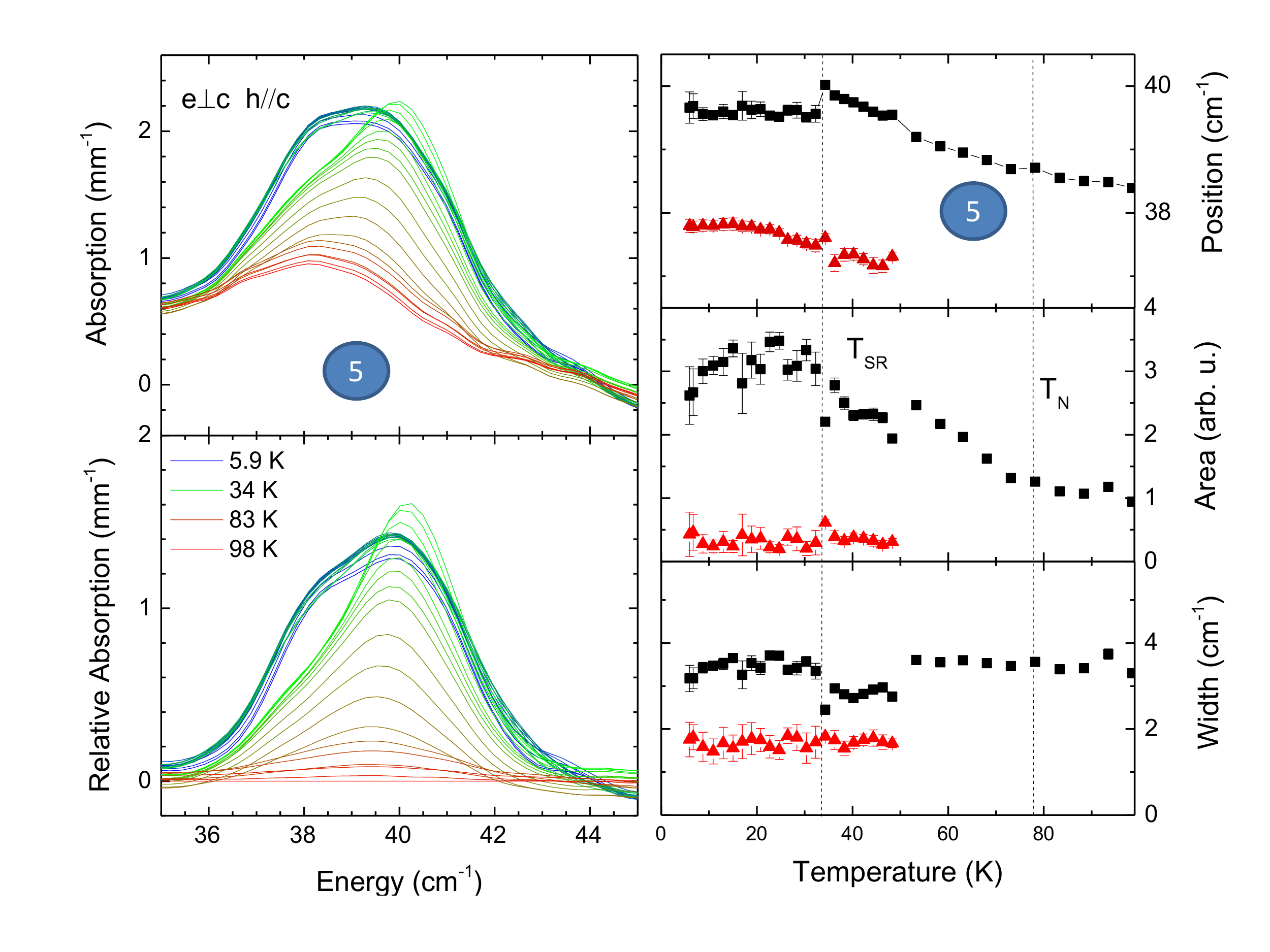}}
 \caption{Details of the THz spectra around band 5 when ${\bf h}\parallel{\bf c}$. Left: absolute and relative absorption measured at different temperatures. Right: results of gaussian fits for the absorption peaks with the position, area and width as a function of temperature. Two peaks are present. The lower energy peak appears smoothly when lowering the temperature and can be fitted below 50 K. The higher energy one, the stronger, is affected at T$_{SR}$ with a small shift in energy and increased width and area.}
 \label{fig6}
 \end{figure}

\begin{figure}
\resizebox{8.5cm}{!}{ \includegraphics{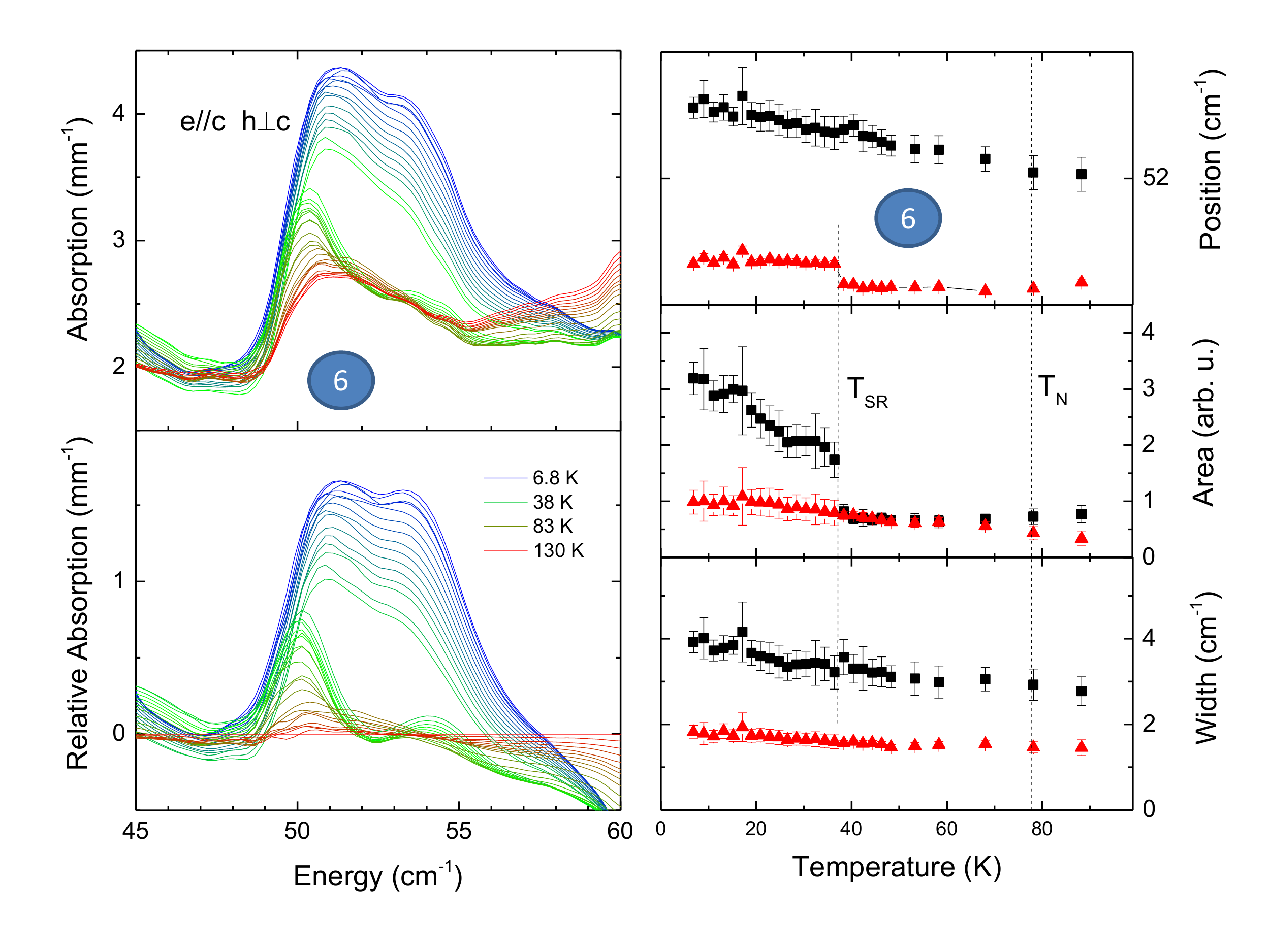}}
 \caption{Details of the THz spectra around band 6 when {\bf h$\perp$c}. Left: absolute and relative absorption measured at different temperatures. Right: results of gaussian fits for the absorption peaks with the position, area and width as a function of temperature. Two peaks are present that are both affected at T$_{SR}$.}
 \label{fig7}
 \end{figure}

A detailed description of the temperature dependence of the spectra are given in Fig.\ref{fig5} to Fig.\ref{fig7}. Six different bands (each possibly including several excitations) are clearly visible together with the Mn magnon. The absorption peaks have been fitted with gaussian profile and their position, spectral weight (area) and lifetime (width) are reported. They all show, at the reorientation transition T$_{SR}$, a slight change of their position in energy and more or less pronounced change of spectral weight. In particular band 1 is observable only below T$_{SR}$ and for {\bf h$\perp$c} (see Fig.\ref{fig3}). Band 2, always present for {\bf h$\perp$c}, is observable only below T$_{SR}$ when ${\bf h}\parallel{\bf c}$ (see Fig.\ref{fig3}). Band 3, always present for {\bf h$\perp$c}, disappears below T$_{SR}$ when ${\bf h}\parallel{\bf c}$, while band 4 has a strong increase below T$_{SR}$  (see Fig.\ref{fig4}). Note that band 3 below T$_{SR}$ might have a shoulder and that band 4 encompasses certainly several modes. Finally, band 6 and maybe also band 5 have a side band that increases below T$_{SR}$ (see Figs.\ref{fig6} and \ref{fig7}). The Mn magnon (M) is also affected by the spin reorientation at T$_{SR}$: an abrupt change, although weak, occurs in its position and line width while its area, which was increasing below T$_N$, starts to decrease and becomes temperature independent below 10 K (see Fig.\ref{fig5}).

A comparison with the neutron data can be made keeping in mind that neutron scattering measures the imaginary part of the dynamical susceptibility that is equal to the THz absorption divided by the energy. Both techniques show remarkable agreement: At  20 K, below T$_{SR}$, the observed five flat bands along $(1, 0, \ell)$) in the neutron data (see Fig.\ref{INS2-v4}) correspond nicely to band 1 - 4 and to the magnon M observed in THz spectroscopy. The band 6 is also visible as a non dispersive branch at low temperature in the neutron data of Fig.\ref{INS1-v4}. At 45 K, above T$_{SR}$, five branches are visible in the neutron spectra, three of them correspond to bands 2-3-4 in THz, while the dispersive magnon M' is not distinguishable in THz, probably because it is below the energy window of the THz spectroscopy or mixed inside bands 3-4 (see Fig.\ref{INS2-v4}). To go further into the analysis of the different magnon/crystal field contributions and their possible hybridization, we propose additional modeling of the complex Mn and Ho dynamics. We first focus on the  crystal field excitations associated to the Ho ions.

\section{CRYSTAL FIELD CALCULATIONS}

\begin{figure}[!h]
\resizebox{8.6cm}{!}{ \includegraphics{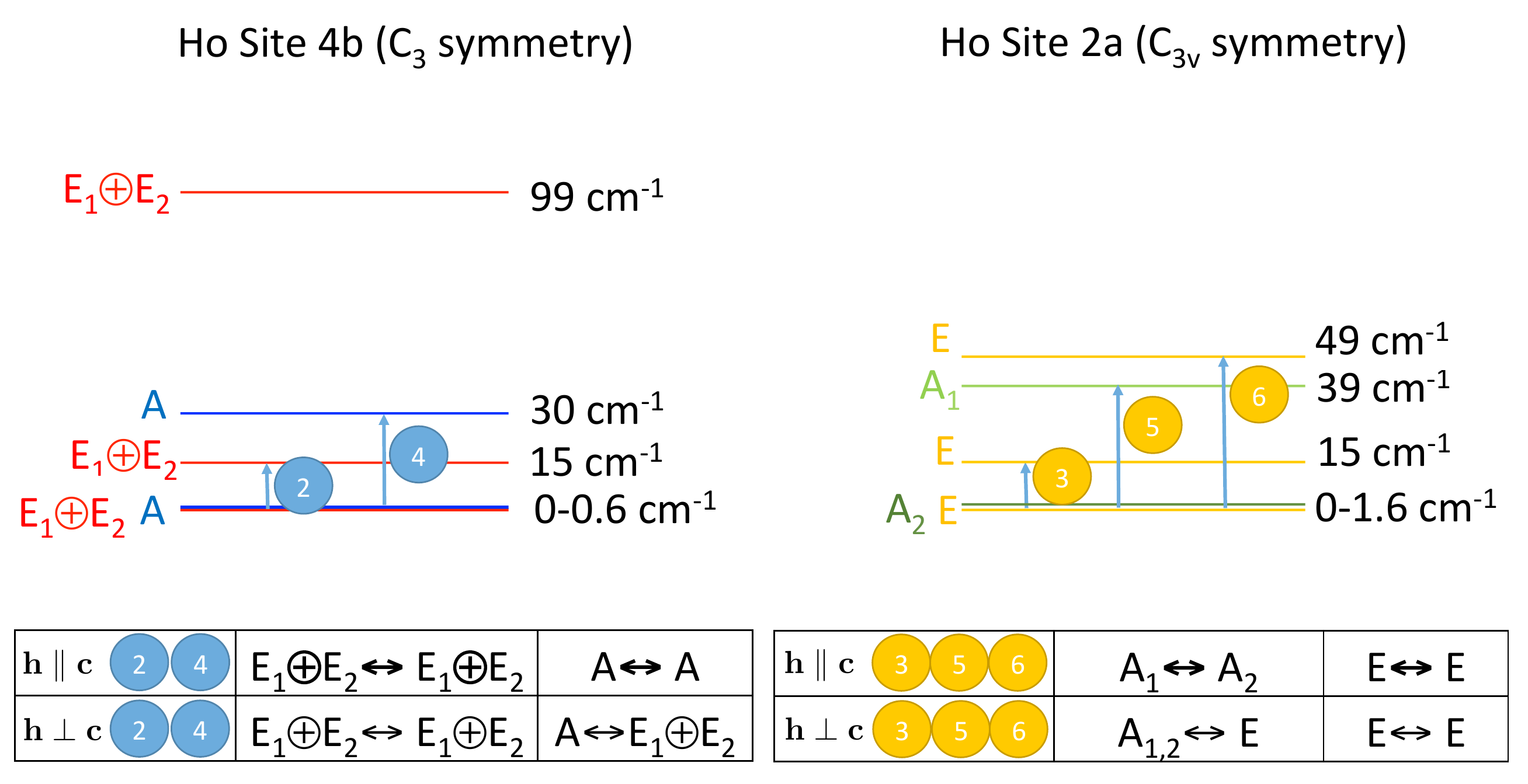}}
 \caption{Crystal field scheme and selection rules of the parameters set 1 for the two different Holmium sites with a tentative matching of the calculated CF transitions with the observed THz bands: Two transitions are expected for site 4b corresponding to band 2 and 3 and  three for site 2a corresponding to band 4, 5 and 6. The selection rules and the symmetry of the levels are indicated below.}
 \label{fig9}
 \end{figure}

Crystal field effects were calculated in a point charge model taking into account the whole oxygen coordination shell of the Ho$^{3+}$ ions (4$f^{10}$) for site 4b in C$_3$ symmetry and site 2a in C$_{3v}$ symmetry \cite{Stevens1952,Hutchings1965,Chaix2014}. The atomic coordinates were obtained from a refinement of neutron diffraction data at 10 K \cite{Fabreges2009}. The associated Hamiltonian for each Ho$^{3+}$ ion in the ground multiplet $J$=8 can be expressed as:
\begin{equation}
\begin{aligned}
H^{CF}_{4b}=A_{2}^{0}O_{2}^{0}+A_{4}^{0}O_{4}^{0}+A_{4}^{3}O_{4}^{3}-(A_{4}^{3})^*O_{4}^{-3}+A_{6}^{0}O_{6}^{0}\\
+A_{6}^{3}O_{6}^{3}-(A_{6}^{3})^*O_{6}^{-3}+A_{6}^{6}O_{6}^{6}+(A_{6}^{6})^*O_{6}^{-6}
\end{aligned}
\label{CF4b}
\end{equation}
\begin{equation}
\begin{aligned}
H^{CF}_{2a}=A_{2}^{0}O_{2}^{0}+A_{4}^{0}O_{4}^{0}+A_{4}^{3}(O_{4}^{3}-O_{4}^{-3})+A_{6}^{0}O_{6}^{0}\\
+A_{6}^{3}(O_{6}^{3}-O_{6}^{-3})+A_{6}^{6}(O_{6}^{6}+O_{6}^{-6})
\end{aligned}
\label{CF2a}
\end{equation}
for sites 4b and 2a, in terms of $O_{n}^{m}$ Racah operators. A global screening factor of 20\% was estimated in order to obtain the transition between the CF levels in the experimental energy range. The CF parameters resulting from this calculation, labeled set 1, are given in Table \ref{tableCF}. The deduced energy scheme and selection rules are given in Fig.\ref{fig9}. For Ho 4b, the energy scheme contains singlet ($A$) and degenerate doublets ($E_{1}\bigoplus E{_2}$). Four different energy levels are calculated below 100 cm$^{-1}$. For Ho 2a, two singlets ($A_1$, $A_2$) and one doublet ($E$) are present with four different energy levels in the same energy range. They are all allowed to occur for any orientation of the THz magnetic field. The calculated absorption is shown in Fig. \ref{fig11}(a) and (b) for both orientations of the THz magnetic field $h$ at 40 K and 6 K respectively. The energy and relative intensities of the highest energy bands 5 and 6 are correctly reproduced. However the agreement between the calculations and the observed complex set of mixed excitations below 30 cm$^{-1}$ is less accurate.

\begin{table}[!htbp]
 \begin{ruledtabular}
 \begin{tabular} {ccc}
set 1& Ho (4b) & Ho (2a) \\
\hline
$A_2^0$  &  0.392332                    & -0.664753   \\
$A_4^0$  &  0.000747                    & -0.000978   \\
$A_4^3$  & -0.037963+i 0.008016 & -0.039276   \\
$A_6^0$  & -0.000776                    & -0.000848   \\
$A_6^3$  &  0.000337-i 0.000075  &  0.000219  \\
$A_6^6$  & -0.000492                    & -0.000531   \\
\hline	
set 2 & Ho (4b) & Ho (2a) \\
\hline
$A_2^0$  &  0.392332                    & -0.531802  \\
$A_4^0$  &  0.000747                    & -0.000978  \\
$A_4^3$  & -0.037963+i 0.008016 & -0.039276  \\
$A_6^0$  & -0.000705                    & -0.000917  \\
$A_6^3$  &   0.000306-i 0.000068  &  0.000237  \\
$A_6^6$  & -0.000448                    & -0.000574   \\ \hline	
\end{tabular}
\end{ruledtabular}
\caption{ CF parameters of sets 1 and 2 for the two Ho sites in the Racah formalism (see text). }
 \label{tableCF}
\end{table}

\begin{figure}
\resizebox{8.3cm}{!}{ \includegraphics{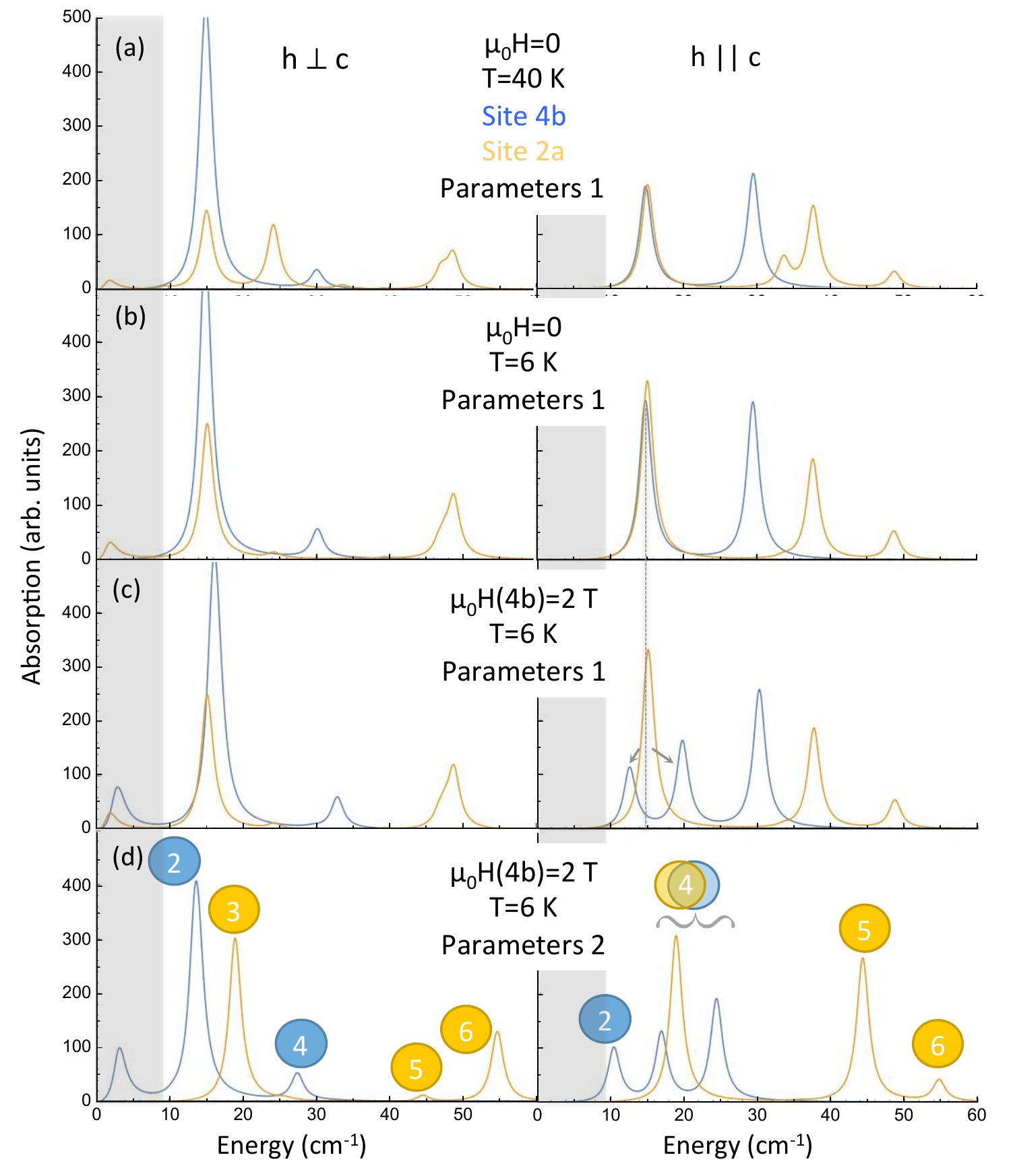}}
 \caption{Calculated CF THz absorption spectra for both orientations of the THz magnetic field $h$ and for Ho sites 4b (blue) and 2a (yellow), using parameters set 1 at 40 K and zero magnetic field (a), at 6 K and zero magnetic field (b) and at 6 K with a 2 T magnetic field parallel to the $c$-axis. The dashed line and the two grey arrows for ${\bf h}\parallel{\bf c}$ indicate the splitting of the lowest energy Ho 4b CF excitation under magnetic field. Panel (d) corresponds to the calculations performed using the parameter set 2 at 6 K and under a magnetic field of 2 T along the $c$-axis with a tentative matching of the calculations with the measured bands of excitations.}
 \label{fig10}
 \end{figure}

To go further, we took into account the coupling of the Ho 4b magnetic moments with the Mn ones expected to occur below T$_{SR}$ (discarding a possible magnetic coupling of the Ho 2a with Mn) \cite{Nandi2008}. This was achieved by including in the calculations the influence of the local magnetic field produced by the Mn moments on the Ho 4b site, similarly to the case of ErMnO$_3$ \cite{Chaix2014}. This field was chosen along the c-axis according to the observed c-direction of the Ho ordered magnetic moments. Its value was adjusted at 2 T so that the ordered Ho 4b magnetic moment extracted from the neutron diffraction data is reproduced at low temperature (see Appendix I). The influence of this magnetic field is to change slightly the energy of some peaks and to produce additional splitting of some CF levels (see Fig.\ref{fig10}b and c for the parameter set 1). We can notice in particular that the lowest energy CF peak corresponding to Ho 4b (in blue) in zero field splits into two peaks under magnetic field. However, experimentally, there are abrupt changes at T$_{SR}$ for all the THz bands, both on their position and their intensity, which cannot be accounted for by the sole effects of temperature and molecular field. This is a strong indication that electric effects with modification of electric charge screening are at play at T$_{SR}$. This is further corroborated by the observation simultaneously of a sharp anomaly in the dielectric constant as well as a jump in the electric polarisation along the c-axis \cite{Hur2009}.

With little additional changes of some CF parameters (set 2 in Table \ref{tableCF}), we can reproduce qualitatively most of the observed features in the THz spectra (see Fig. \ref{fig11}d for T=6 K and a field of 2 T). In particular, bands 2, 3, 4, 5 and 6 are predicted at energies closer to the measured ones, although there are still some discrepancies with the calculated intensities. The presence of peak 3 above T$_{SR}$ for ${\bf h}\parallel{\bf c}$ and its disappearance below T$_{SR}$ while a new peak labeled 2 appears at lower energy could be due to the field splitting ascribed to the Ho-Mn coupling. Band 4 for ${\bf h}\parallel{\bf c}$ below T$_{SR}$ could then result from a mixture of several peaks including the higher energy mode from the split peak (see horizontal bracket in Fig. \ref{fig11}d).

Strikingly, for ${\bf h}\perp{\bf c}$,  band 1 appearing at T$_{SR}$ in the measurements is missing in the calculations and therefore cannot be accounted for by the effect of a magnetic field. This discrepancy between calculations and experiment is another indication that hybridization is at play between Ho CF excitations and the lowest energy Mn magnon (M'). 

\section{Discussion}

Our analysis and calculations for the Mn magnetic order and associated spin waves, using the spin Hamiltonian of eq.\ref{H3d}, can account for most of the experimental data. Mainly, two spin branches M and M' are  observed with INS that are correctly calculated (Fig.\ref{INS1-v4} and \ref{INS2-v4}). In particular, the spin reorientation at T$_{SR}$ is clearly seen as a change in the spin wave dispersion along ${\bf c^*}$ for the  branch M' and is correctly captured when the change of Mn spin configuration from $\Gamma_4$ to $\Gamma_3$ is introduced. The second  branch M is also observed with THz spectroscopy (Fig.\ref{fig5}), in agreement with the model.
As for the non dispersive Ho CF excitations observed with INS and THz spectroscopy, our calculations using the CF hamiltonian  described by eq.\ref{CF4b} and \ref{CF2a} is able to reproduce the experimental data qualitatively when two ingredients are included: first, a change in charge screening at T$_{SR}$  that explains the observed shifts and changes in intensity of the whole CF spectrum, and second, a molecular field along the c-axis produced by the Mn order. Similarly,  a molecular  field produced by the Ho 4b order  may be introduced to account for the increased anisotropy gap in the Mn spin wave branch M'. However our calculations do not agree quantitatively with the whole experimental data: they cannot describe properly how the Mn spin branch M' become non dispersive along $(1, 0, \ell)$ and seems to merge into Ho CF excitations below 20 K.  Moreover, they cannot explain the THz band 1 that appears at T$_{SR}$ and is locked at 3 cm$^{-1}$ below band 2 (Fig.\ref{fig3}). Additional ingredients should be at play there, that couple the 3d and 4f magnetic species dynamically.

It is possible for  Mn Magnons and Ho crystal field excitations to hybridize, if they are  close in energy,  which is the case for the Ho manganite where 6 different Ho crystal field excitations lie in the energy range of the Mn spin waves (see Fig.\ref{INS2-V4}). They also should not be forbidden by symmetry. This can be checked using  a group representation analysis for both the Mn spin waves and the Ho crystal field excitations : if they contain at least one common irreducible representation, the coupling is allowed. The details are given in appendix II. According to  \cite{Sikora1988}, for the Mn magnetic orders with zero propagation vector corresponding to the four different one dimensionnal $\Gamma_i$, the spin wave representation  may be decomposed into $\tau = A_{2}\bigoplus B_{1}\bigoplus E{_1}\bigoplus E{_2}$ where we use the irreducible representations of the  C$_{6v}$ (6mm) point group. For Ho 4b and 2a crystal field excitations, their decompositions are already given in Fig.\ref{fig9}, in the local point groups associated to the two different sites: C$_{3}$ (3) for Ho 4b, and C$_{3v}$ (3m) for Ho 2a. Their energy scheme contains singlets ($A$ for Ho 4b, $A_1$, $A_2$ for Ho 2a) and doublets ($E_1$, $E_2$ for Ho 4b, $E$ for Ho 2a). To infer compatibility with those of the Mn magnons, we must consider their decomposition into the irreducible representations of the C$_{6v}$ point group. For Ho 4b,  singlets $A$ are decomposed into \{$A_{1},A_{2},B_{1},B_{2}$\} and  degenerate doublets ($E_{1}\bigoplus E{_2}$ are decomposed into \{$E_{1},E{_2}$\}. As for the Ho 2a,  singlets $A_{1}$ and $A_{2}$  are decomposed into \{$A_{1}, B_{2}$\} and \{$A_{2}, B_{1}$\} respectively, and doublets $E$ into \{$E_{1},E{_2}$\} as for Ho 4b. It can be easily seen that they all have in common at least one irreducible representation contained in the Mn spin wave representation except for Ho 2a $A_{1}$ singlet that is calculated around 40-45 cm$^{-1}$ and corresponds to band 5.

\begin{figure}[!h]
\includegraphics[width=8.8cm]{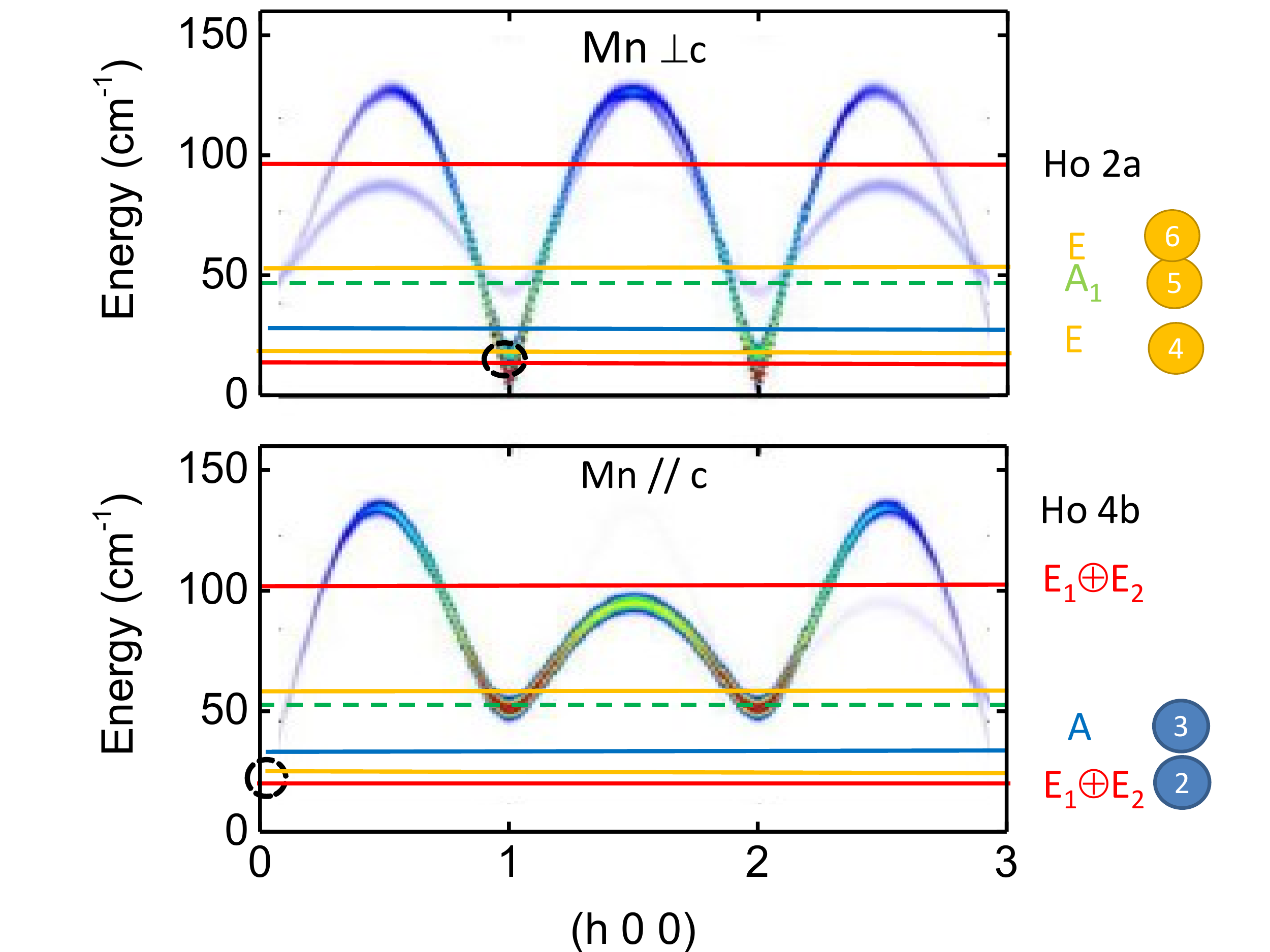}
\caption{(a) Mn spin wave calculations along $(h, 0, 0)$ for spin fluctuations within the (a, b) plane (upper panel) and along the c-axis (lower panel). The calculated CF levels are added for the Ho 4b site (red and  blue lines) as well as the Ho 2a site (yellow and green lines). The coupling of these CF excitations with the Mn magnons is always allowed by symmetry for all but for the dashed one.}
\label{figSWCF}
\end{figure}

 We conclude then that, in the energy range relevant for the experimental data reported in this study, all the CF excitations, except one (band 5 associated to Ho 2a), are allowed to couple to Mn spin waves.  Looking back at neutron experiments and the spin wave calculations reproduced in Fig.\ref{figSWCF}, it can be seen that, when the temperature is lowered close to T$_{SR}$, the magnon M' with small gaps at the zone centers (1,0,0) and (1,0,1) become close in energy to Ho (4b) CF excitations (bands 2 and 3-4), with different spectral weight for the magnons depending on the magnetic order $\Gamma_{1,3}$ or $\Gamma_{2,4}$. A coupling mechanism may be at play here that contributes to the spin reorientation process and the lost of dispersion for M' along ${\bf c^*}$.

It is worth noting that Ho and Mn ordered magnetic moments below T$_{SR}$ are orthogonal. This suggests that, in the Mn - Ho coupling Hamiltonian of eq.\ref{Hcc}, the coupling tensor $g$ shows non diagonal terms, as it is the case for instance for the pseudo dipolar, the Dzyaloshinkii-Moriya interaction \cite{Fabreges2010} or the highest order trigonal anisotropy \cite{Plumer2010}. Furthermore, the 2 T effective magnetic field experienced by the 4b Ho suggests a coupling strength $g$ $\sim$ 0.1 K. In this context and given the neutron scattering data and the CF analysis detailed above, we may to consider a possible dynamical coupling between i) in plane Mn spin wave modes with Ho CF transitions polarized along the c-axis and ii) out of plane Mn spin wave modes with Ho CF transitions polarized in the (a,b) plane. In this view, it is tempting to assign band 1 to the Mn spin wave hybridized with band 2 (and also likely with band 3). In this scenario, the in plane character of the CF transitions would be transferred to the hybrid mode which would now appear in the {\bf h$\perp$c} channel as observed experimentally in the THz measurements. This hybridisation occurs only below  T$_{SR}$ when the spectral weight of these excitations is enhanced and their energy matches. The energy difference between band 1 and 2, around 3 cm$^{-1}$ experimentally, is in agreement with the coupling strength $g$ when all the Ho-Mn links are considered (12 of them).

\section{CONCLUSION}

To summarize, we have reexamined the THz dynamics present in the multiferroic compound \homn. The peculiarity of this Ho compound in the hexagonal manganites is the occurrence of a spin reorientation below T$_{SR}$=37 K. Remarkably, several Ho CF excitations are present in the same energy range as the Mn spin waves. Thanks to the complementary neutron and electromagnetic wave probes, we have identified five Ho 4b and Ho 2a CF excitations and shown how Mn spin waves as well as Ho CF are affected by the reorientation process at T$_{SR}$. Using spin wave and CF calculations, we are able to reproduce qualitatively most of the observed features. At T$_{SR}$, the Mn spin configuration switches from the high temperature $\Gamma_4$ configuration to $\Gamma_3$. A change in the spectral weight is observed along the $(1, 0, \ell)$ direction that is correctly calculated in a simple hamiltonian for the Mn win waves. At the same time, changes occur in the CF excitations that can be accounted for when coupling to the lattice and surrounding oxygen charges are introduced. Then, Ho CF excitations start to match in energy the Mn branch at different Brillouin zone centers. Further inspection of the neutron and THz data points towards additional Ho-Mn coupling mechanisms that show up below T$_{SR}$ in two different ways : first as an additional excitation observed in THz spectroscopy that is locked 3 cm$^{-1}$ below a CF excitation and secondly the progressive disappearance of the Mn dispersion along ${\bf c^*}$.
Our detailed study on \homn\ gives finally new insight on the very rich rare earth - transition metal dynamical couplings that show up in the THz range.

\acknowledgments This work was financially supported by the ANR-13-BS04-0013-01.

\appendix
\section{APPENDIX I}

\begin{figure}[!h]
\resizebox{7cm}{!}{ \includegraphics{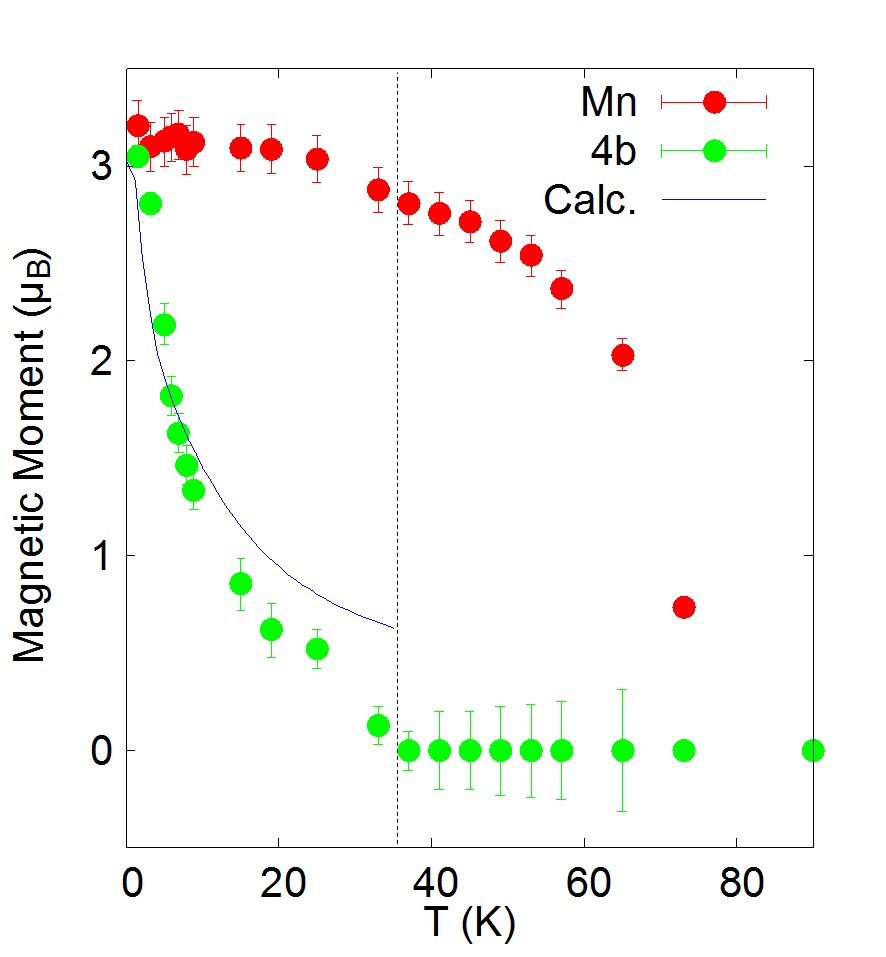}}
 \caption{Temperature dependence of the ordered magnetic moments Ho 4b extracted from the refinement of the neutron diffraction data (black dots) compared to the calculated magnetic moment (blue line) induced by a contant magnetic field of 2 T due to the coupling with the Mn magnetism. The red stars show the Mn ordered moment also extracted from the neutron diffraction data.}
 \label{fig11}
 \end{figure}

In the Ho CF calculations, we introduce a constant magnetic field  to simulate  the effect of the Ho-Mn coupling. We used the measured Ho 4b and Mn ordered moment extracted from neutron diffraction data \cite{Fabreges2010}: Ho 4b moments are observed along the c-axis; a molecular field of 2 T along this direction is able to polarize Ho spins with the correct magnitude and temperature dependence at least below 10 K (see Fig.\ref{fig11}). The Mn moment varies only slightly below T$_{SR}$ thus validating the assumption of a constant molecular field.

\section{APPENDIX II}

For Mn Magnons and Ho crystal field excitations to hybridize, there may be symmetry requirements that can be derived through a group representation analysis for the Mn spin waves and the Ho crystal field excitations. The Mn order with zero propagation vector is described by the isomorphic space group C$_{6v}^3$. The Mn spin wave representations can be reduced into representations of the associated point group C$_{6v}$ (6mm). This group contains the lower symmetric C$_{3}$ (3) and C$_{3v}$ (3m)that are best describing the Ho CF shemes. In the present analysis, to look at possible coupling mechanisms with the Mn spin waves,  we will rewrite the CF representations in the more symmetric C$_{6v}$.  Its character table is given in table \ref{Tbl:6mm}.

 The symmetry of the different Mn spin waves have been reported in \cite{Sikora1988} with Kovalev notation. We will rather use Mulliken notation. To simplify calculations, the spin wave analysis has been performed in a more symmetric space group than C$_{6v}^3$, namely D$_{6h}^4$, which means neglecting some oxygen displacements and therefore some weak magnetic anisotropies. From \cite{Sikora1988}, it follows that the spin wave representation  can be decomposed into A$_{2g} \bigoplus B_{1g} \bigoplus E_{1g} \bigoplus E_{2g}$. Going back to the lower symmetric C$_{6v}^3$ space group, gerade und ungerade modes are no longer relevant so that the spin wave decomposition becomes A$_{2}\bigoplus B_{1}\bigoplus E_{1}\bigoplus E{_2}$. When spin anisotropies are introduced, the degeneracy of the low  energy branches is lifted as can be seen in our spin wave calculations  (Fig.\ref{INS1-v4}), dictating the use of magnetic Shubnikov groups.

For Ho CF schemes,  The character tables for the associated point groups C$_{3}$ (3) and C$_{3v}$ (3m) are reported in table \ref{Tbl:3} and \ref{Tbl:3v}. Their correlations to the more symmetric C$_{6v}$ group that we will use are given in table \ref{Tbl: Cor C3}. For Ho 4b, in its local C$_{3}$ symmetry, two different kind of symmetry are present for the crystal field states: singlets are described by the irreducible representation $A$ and degenerate doublets described by E$_{1}\bigoplus E{_2}$. These irreducible representations induce, in the higher symmetric C$_{6v}$ group, states that are decomposed into linear combinations of \{A$_{1},A_{2},B_{1},B_{2}$\} for the singlets and \{E$_{1},E{_2}$\} for the degenerate doublets. For the Ho 2a, in the local C$_{3v}$ symmetry, three different symmetries are present: singlets A$_{1}$ and A$_{2}$, and  doublets E. They induce, in  C$_{6v}$ symmetry, states that are decomposed into \{A$_{1},B_{2}$\} and \{A${_2},B_{1}$\} respectively for the singlets and \{E$_{1},E_{2}$\} for the doublets.

From these considerations, we can now look at possible coupling between the Mn spin waves and the Ho 4b and 2a crystal field excitations that are allowed by symmetry, that is if their representations contain at least one common irreducible representation. This is the case fot all CF excitations except for the  Ho 2a  A$_1$ singlets.

\begin{table}[!htbp]
 \caption{ $C_{6v}$ (6mm) character table }
 \label{Tbl:6mm}
 \begin{ruledtabular}
 \begin{tabular} {ccccccccc}
$C_{6V}$ & 6mm & 1 & 2 & 3 & 6 & $m_d$ & $m_v $& functions \\
\hline
Multiplicity & - & 1 & 1 & 2 & 2 & 3 & 3 & . \\
$A_{1}$  & $\Gamma_{1}$ & 1 & 1 & 1& 1 & 1 & 1 & z, $x^{2}+y^{2}$, $z^{2}$  \\
$A_{2}$  & $\Gamma_{2}$ & 1 & 1 & 1& 1 & -1 & -1 & $J_{z}$  \\
$B_{1}$  & $\Gamma_{3}$ & 1 & -1 & 1& -1 & 1 & -1 & .  \\
$B_{2}$  & $\Gamma_{4}$ & 1 & -1 & 1& -1 & -1 & 1 &  .  \\
$E_{2}$  & $\Gamma_{5}$ & 2 & 2 & -1& -1 & 0 & 0 & z, $x^{2}-y^{2}$, xy  \\
$E_{1}$  & $\Gamma_{6}$ & 2 & -2 & -1& 1 & 0 & 0 & x, y, xz, yz $J_{x}$, $J_{y}$  \\ \hline	
\end{tabular}
\end{ruledtabular}
\end{table}

\begin{table}[!htbp]
 \caption{ $C_{3}$ (3) character table }
 \label{Tbl:3}
 \begin{ruledtabular}
 \begin{tabular} {cccccc}
  $C_{3}$& 3 & 1 & $3^{+}$ & $3^{-}$ & functions \\
\hline
  $A$  & $\Gamma_{1}$ & 1 & 1 & 1&  z, $x^{2}+y^{2}$, $J_{z}$ \\
  &   &   &  &  &  \\
  $E_{2}$  & $\Gamma_{2}$ & 1 & $\omega^{2}$ & $\omega$& \\
 &   &   &  &  & x, y, xz, yz, $x^{2}-y^{2}$, xy,$J_{x}$ ,$J_{y}$ \\
  $E_{1}$  & $\Gamma_{3}$ & 1 & $\omega $& $\omega^{2}$&  \\ \hline
 	
\end{tabular}
\end{ruledtabular}
\end{table}

\begin{table}[!htbp]
 \caption{ $C_{3v}$ (3) character table }
 \label{Tbl:3v}
 \begin{ruledtabular}
 \begin{tabular} {cccccc}
  $C_{3v}$& 3m & 1 & 3 & m & functions \\
\hline
  Multiplicity & - & 1 & 2 & 3\\
  $A_{1}$  & $\Gamma_{1}$ & 1 & 1 & 1&  z, $x^{2}+y^{2}$, $z^{2}$ \\
  $A_{2}$  & $\Gamma_{2}$ & 1 & 1 & -1 &$J_{z}$\\
  E & $\Gamma_{3}$ & 2 & -1 & 0 & x, y, xz, yz, $x^{2}-y^{2}$, xy,$J_{x}$ ,$J_{y}$ \\ \hline
 	
\end{tabular}
\end{ruledtabular}
\end{table}

All the tables reproduced here where established thanks to \cite{Aroyo2006}.

\begin{table}[!htbp]
 \caption{Correlations between the global frame point group $C_{6v}$ (6mm) and the two local frame point groups $C_{3}$ (3) and $C_{3v}$ (3m) }
 \label{Tbl: Cor C3}
 \begin{ruledtabular}
 \begin{tabular} {ccc}
  $C_{6v}$ & $C_{3}$ & $C_{3v}$ \\
\hline
  6mm  & 3 & 3m \\
  $A_{1}$  & A & $A_{1}$\\
$A_{2}$  & A & $A_{2}$\\
$B_{1}$  & A & $A_{2}$\\
$B_{2}$  & A & $A_{1}$\\
$E_{2}$  & $E^{1},E^{2}$ & E\\
$E_{1}$  & $E^{1},E^{2}$ &E \\ \hline	
\end{tabular}
\end{ruledtabular}
\end{table}



\begin{thebibliography}{00}

\bibitem{Pimenov2006} A. Pimenov, A. A. Mukhin, V. Yu. Ivanov, V. D. Travkin, A. M. Balbashov, A. Loidl, Nat. Phys. {\bf 2} 97 (2006).
\bibitem{Sushkov2007}A. B. Sushkov, R. V. Aguilar, S. Park, S-W. Cheong, and H. D. Drew, Phys. Rev. Lett. {\bf 98}, 027202 (2007).
\bibitem{Kida2009}N. Kida, D. Okuyama, S. Ishiwata, Y. Taguchi, R. Shimano, K. Iwasa, T. Arima, and Y. Tokura Phys. Rev. B {\bf 80}, 220406(R)(2009).
\bibitem{Jones2014}S.P.P. Jones, S.M. Gaw, K.I. Doig, D. Prabhakaran, E.M. H\'{e}troy Wheeler, A.T. Boothroyd, and J. Lloyd-Hughes, Nature Com. {\bf 5},  3787 (2014).
\bibitem{Katsura2007}H. Katsura, A.V. Balatsky,and N. Nagaosa, Phys. Rev. Lett. 98, 027203 (2007).
\bibitem{Mochizuki2010b} M. Mochizuki, N. Furukawa,and N. Nagaosa, Phys. Rev. Lett. {\bf 105}, 037205 (2010).
\bibitem{Mochizuki2010} M. Mochizuki, N. Furukawa,and N. Nagaosa, Phys. Rev. Lett. {\bf 104}, 177206 (2010).
\bibitem{Chaix2013} L. Chaix, S. de Brion, F. L\'evy-Bertrand, V. Simonet, R. Ballou, B. Canals, P. Lejay, J. B. Brubach, G. Creff, F. Willaert, P. Roy, and A. Cano, Phys. Rev. Lett. {\bf 110}, 157208 (2013).
\bibitem{Sirenko2008} A. A. Sirenko, S. M. OMalley, K. H. Ahn, S. Park, G. L. Carr, and S.-W. Cheong, Phys. Rev. B {\bf 78}, 174405 (2008).
\bibitem{Kang2010} T. D. Kang, E. Standard, K. H. Ahn, A. A. Sirenko, G. L. Carr, S. Park, Y. J. Choi, M. Ramazanoglu, V. Kiryukhin, and S.-W. Cheong, Phys. Rev. B 82, 014414 (2010).
\bibitem{Chaix2014} L. Chaix, S. de Brion, S. Petit, R. Ballou, L.-P. Regnault, J. Ollivier, J.-B. Brubach, P. Roy, J. Debray, P. Lejay, A. Cano, E. Ressouche and V. Simonet, Phys. Rev. Lett. {\bf 112}, 137201 (2014).
\bibitem{Pailhes2009} S. Pailh\`es, X. Fabr\`eges, L. P. R\'egnault, L. Pinsard-Godart, I. Mirebeau, F. Moussa, M. Hennion, and S. Petit, Phys. Rev. B {\bf 79}, 134409 (2009).
\bibitem{Toth2016}S. T\'{o}th, B. Wehinger, K. Rolfs, T. Birol, U. Stuhr, H. Takatsu, K.Kimura,T. Kimura, H.M. R{\o}nnow, and C.R\"{u}egg, Nature Com. {\bf Y},13547 (2016).
\bibitem{Aupiais2018} I. Aupiais, M. Mochizuki, H. Sakata, R. Grasset, Y. Gallais, A. sacuto, and M. Cazayous, npj Quantum Materials {\bf 3}, 60 (2018).
\bibitem{Bertaut1963} E. F. Bertaut, F. Forrat, and P. Fang, Compt. Rend. Acad. Sci.(Paris) {\bf 256}, 1958 (1963).
\bibitem{Katsufuj2002} T. Katsufuji, M. Masaki, A. Machida, M. Moritomo, K. Kato, E. Nishibori, M. Takata, M. Sakata, K. Ohoyama, K. Kitazawa, and H. Takagi, Phys. Rev. B {\bf 66}, 134434 (2002).
\bibitem{Cheong2007} S.W. Cheong, M. Mostovoy, Nature materials, {\bf 6}, 13 (2007).
\bibitem{Sim2016} H. Sim, J. Oh, J. Jeong, M. Duc Le, and J.-G. Park, Acta Crystallogr., Sect. B {\bf 72}, 3 (2016).
\bibitem{Fabreges2009} X. Fabr\`eges, S. Petit, I. Mirebeau, S. Pailh\`es, L. Pinsard, A. Forget, M. T. Fernandez-Diaz, and F. Porcher, Phys. Rev. Lett. {\bf 103}, 067204 (2009).
\bibitem{Lee2008} S. Lee, A. Pirogov, M. Kang, K.-H. Jang, M. Yonemura, T. Kamiyama, S.-W. Cheong, F. Gozzo, N. Shin, H. Kimura, Y. Noda and J.-G. Park, Nature {\bf 451}, 805 (2008).
\bibitem{Solovyev2014} I. V. Solovyev, and S.A. Nikolaev, Phys. Rev. B {\bf 90}, 184425 (2014).
\bibitem{Fabreges2008} X. Fabr\`eges, I. Mirebeau, P. Bonville, S. Petit, G. Lebras-Jasmin, A. Forget, G. Andr\'e, and S. Pailh\`es, Phys. Rev. B {\bf 78}, 214422 (2008).
\bibitem{Chattopadhyay2018} S. Chattopadhyay, V. Simonet, V. Skumryev, A. A. Mukhin, V.Y. Ivanov, M. I. Aroyo, D. Z. Dimitrov, M. Gospodinov, and E. Ressouche, Phys. Rev. B {\bf 98}, 134413 (2018).
\bibitem{Lottermoser2004} T. Lottermoser, T. Lonkai, U. Amann, D. Hohlwein, J. Ihringer, and M. Fiebig, Nature {\bf 430}, 541 (2004).
\bibitem{Munoz2001} A. Munoz, J. A. Alonso, M. J. Mart\'inez-Lope, M. T. Cas\'ais, J. L. Mart\'inez, and M. T. Fern\'andez-Diaz, Chem. Mater. 1{\bf 3}, 1497 (2001).
\bibitem{Fiebig2002} M. Fiebig, C. Degenhardt, R. V. Pisarev, J. App. Phys. {\bf 91}, 8867 (2002).
\bibitem{Brown2006} P. J. Brown and T. Chatterji, J. Phys. Condens. Matter {\bf 18}, 10085 (2006).
\bibitem{Brown2008} P. J. Brown and T. Chatterji, Phys. Rev. B {\bf 77}, 104407 (2008).
\bibitem{Nandi2008} S. Nandi, A. Kreyssig, L. Tan, J. W. Kim, J. Q. Yan, J. C. Lang, D. Haskel, R. J. McQueeney, and A. I. Goldman, Phys. Rev. Lett. {\bf 100}, 217201 (2008).
\bibitem{Hur2009} N. Hur, I.K. Jeong,M .F. Hundley, S.B. Kim and S.-W. Cheong, Phys. Rev. B {\bf 79}, 134120 (2009).
\bibitem{Talbayev2008} D. Talbayev, A.D. LaForge, S.A. Trugman, N. Hur, A.J. Taylor, R.D. Averitt and D.N. Basov, Phys. Rev.  {\bf 101}, 247601 (2008).
\bibitem{Laurita2017} N.J. Laurita, Yi Luo, Rongwei Hu,, Meixia Wu, S.W. Cheong, O. Tchernyshyov and N.P. Armitage, Phys. Rev. Lett. {\bf 119}, 227601 (2017).
\bibitem{Sato2003} T. J. Sato, S. -H. Lee, T. Katsufuji, M. Masaki, S. Park, J. R. D. Copley, and H. Takagi, Phys. Rev. B {\bf 68}, 014432 (2003).
\bibitem{Park2003} J. Park, J.-G. Park, Gun Sang Jeon, Han-Yong Choi, Changhee Lee, W. Jo, R. Bewley, K. A. McEwen, and T. G. Perring, Phys. Rev. B {\bf 68}, 104426 (2003).
\bibitem{Vajk2005} O. P. Vajk, M. Kenzelmann, J. W. Lynn, S. B. Kim, and S.-W. Cheong, Phys. Rev. Lett. {\bf 94}, 087601 (2005).
\bibitem{Petit2007} S. Petit, F. Moussa, M. Hennion, S. Pailh\`es, L. Pinsard-Gaudart, and A. Ivanov, Phys. Rev. Lett. {\bf 99}, 266604 (2007).
\bibitem{Kim2018} T. Kim, J. Leiner, K. Park, J. Oh, H. Sim, K. Iida, K. Kamazawa, and J.-G. Park, Phys. Rev. B {\bf 97}, 201113(R) (2018).
\bibitem{SpinWave} http://www-llb.cea.fr/logicielsllb/SpinWave/SW.html
\bibitem{Fabreges2010} X. Fabr\`eges, PhD thesis, Universit\'{e} Paris Sud -Paris XI (2010)
\bibitem{Bowlan2016} P. Bowlan, S.A. Trugman, J. Bowlan, J.-X. Zhu, N.J. Hur, A.J. Taylor, D.A. Yarotski and R.P. Prasankumar, Phys. Rev. B {\bf 94}, 100404(R) (2016).
\bibitem{Stevens1952} K. W. H. Stevens, Proc. Roy. Soc. (London) A65, 209 (1952).
\bibitem{Hutchings1965} M. T. Hutchings, Solid State Physics: Advances in Research and Applications, ed. F.Seitz and B.Turnbull (New York: Academic) 16, 227 (1965).
\bibitem{Sikora1988} W. Sikora, O.V. Gurin and V.N. Syromyatnikov, Journal of Magnetism and Magnetic Materials {\bf 71}, 225 (1988).
\bibitem{Plumer2010} S.G. Condran and M.L. PLumer, J. Phys.: Condens. Matter {\bf 22}, 162201 (2010).
\bibitem{Aroyo2006} M. I. Aroyo, A. Kirov, C. Capillas, J. M. Perez-Mato and H. Wondratschek, Acta Cryst. {\bf A62}, 115-128 (2006).





\end{thebibliography}
\end{document}